# Single-photon circularly polarized single-mode vortex beams

Xujing Liu[1,2,3], Yinhui Kan[2,3], Shailesh Kumar[2], Danylo Komisar[2], Changying Zhao[1], Sergey I. Bozhevolnyi[2]*

**Generation of single photons carrying spin and orbital angular momenta (SAM and OAM) opens enticing perspectives for exploiting multiple degrees of freedom for high-dimensional quantum systems. However, on-chip generation of single photons encoded with single-mode SAM-OAM states has been a major challenge. Here, by utilizing carefully designed anisotropic nanodimers fabricated atop a substrate, supporting surface plasmon polariton (SPP) propagation, and accurately positioned around a quantum emitter (QE), we enable nonradiative QE-SPP coupling and the SPP outcoupling into free-space propagating radiation featuring the designed SAM and OAM. We demonstrate on-chip room-temperature generation of well-collimated (divergence < 7.5º) circularly polarized (chirality > 0.97) single-mode vortex beams with different topological charges ($\ell$ = 0, 1, and 2) and high single-photon purity, $g^{(2)}(0)$ < 0.15. The developed approach can straightforwardly be extended to produce multiple, differently polarized, single-mode single-photon radiation channels, and enable thereby realization of high-dimensional quantum sources for advanced quantum photonic technologies.**

Structuring single-photon emission and engineering photon states triggers breakthroughs in fundamental physics[1,2] and underpins a broad range of modern quantum technologies[3-6]. One of the forefront research fields is chiral quantum optics making use of unique degrees of freedom associated with photon spin and orbital angular momenta, SAM[7-9] and OAM[10-12]. The superpositions of right- and left-circular SAM states ($|R\rangle$ and $|L\rangle$) and OAM states ($\pm\ell$) encoded in single photons can advantageously be exploited for advanced applications in high-dimensional quantum information processing[13-17]. Chiral (i.e., SAM encoded) photon emission has been realized at low temperatures by applying a strong magnetic field[18,19] and then splitting the excitation transitions in specially engineered nanostructures, e.g., a photonic crystal waveguide[19]. The low-temperature generation of single photons with different SAM-OAM superposition states has also been demonstrated using QEs coupled to microring resonators[20,21]. Although the requirement of low-temperature operation restricts eventual practical applications, the reported configurations suggest possibilities for on-chip generation of single photons encoded with SAM-OAM states by coupling QEs with individual nanostructures, an approach which is different from using spontaneous parametric down-conversion in bulk non-linear crystals[22,23] or external mode transformation[24-26], all suffering from low photon generation efficiency, low scalability and complicated transformation process.

[1]Institute of Engineering Thermophysics, Shanghai Jiao Tong University, Shanghai, 200240, China. [2]Centre for Nano Optics, University of Southern Denmark, DK-5230 Odense M, Denmark. [3]These authors contributed equally: Xujing Liu, Yinhui Kan.
*e-mail: seib@mci.sdu.dk

Metasurfaces, representing extended surface nanostructures, exhibit unique capabilities for arbitrarily shaping electromagnetic fields, becoming a staple in classical optics[27-32]. Very recently, metasurfaces have been introduced to empower single-photon sources[33-37], opening thereby design pathways unattainable for configurations that involve QEs coupled to individual nanostructures. The "meta-atom" approach, in which a QE is efficiently and nonradiatively coupled to SPP modes that are in turn outcoupled to free-space propagating radiation by QE-encircling dielectric nanoridges[38], is especially attractive due to its efficiency[39], versatility[34,36], and the possibility for room-temperature generation of circularly polarized single photons[38,40]. A careful analysis of the underlying physics involved in the photon generation by such a meta-atom revealed a daunting fundamental challenge: conventional configurations of QE-encircling nanoridges, such as constant-width[34,39] and width-varying[38] concentric nanoridges or Archimedean spiral gratings[40], can efficiently generate only radially polarized photons[36]. Generation of purely circularly right or left polarized (RCP or LCP) photons is *impossible* with spiral or concentric nanoridges, because the SPP outcoupling by these ridges does not change the (radially oriented) SPP polarization. As a result, any spiral or concentric grating would always generate, simultaneously and with equal efficiency, RCP and LCP beams carrying the same total angular momentum (determined by the number of spirals used), i.e., with their OAM (topological charges) being different by two[36,40]. Similar fundamental limitations are also found in other configurations, rendering single-photon generation of single-mode SAM-OAM vortex beams extremely challenging if not impossible[41].

In this work, we propose the general approach that enables *independent* manipulation of the single-photon SAM ($|\Psi\rangle_{SAM}$) and OAM ($|\Psi\rangle_{OAM}$) states by using dimers of orthogonal nanorods (DONRs), generating high purity $|\Psi\rangle_{SAM}$, arranged in metasurface arrays controlling $|\Psi\rangle_{OAM}$. Using nanodiamonds (NDs) containing nitrogen-vacancy (NV) centres positioned precisely within dielectric DONR-based metasurfaces atop silver (Ag) films, we demonstrate a set of room-temperature on-chip single-photon sources of well-collimated circularly polarized single-mode vortex beams. The DONRs are carefully designed to efficiently convert SPP modes (by outcoupling) to free-space propagating RCP photons with near perfect purity (98% for $S_3 > 0.9$) and with the negligibly small (< 0.3%) spatial overlap between RCP and residual LCP photons. Furthermore, by arranging the DONRs along different (spiral and concentric) trajectories, high-purity SAM states are superimposed with OAM states corresponding to different phase-front helicities (topological charges): $\ell = 0, 1$, and 2. The generated single-photon RCP single-mode vortex beams exhibit a low divergence (< 7.5º) as well as robust characteristics with respect to small QE misalignments (± 50 nm).

## Results

**Design of SAM-OAM photon sources.** The general design concept is based on devising individual nanostructures that would outcouple (radially polarized) SPP waves into circularly polarized free space propagating radiation and aligning these nanostructures along trajectories chosen according to the desired OAM. One could opt for resonant plasmonic (metal) nanorods, whose perpendicular dipole moments scatter with the $\pi/2$ phase difference[42,43], but the inevitable photon absorption by metal nanorods would jeopardise the single-photon generation efficiency. We suggest using instead pairs of orthogonal dielectric nanorods, i.e., DONRs, whose separation along the SPP propagation ensures the proper phase difference between their main dipole moments. Following this concept, we place ND-NVs on a Ag substrate covered by a silica ($SiO_2$) spacer and arrange the designed DONRs made of hydrogen silsesquioxane (HSQ) along the OAM-corresponding trajectories (Fig. 1a). Note that the ND-NVs are considered being excited with a tightly focused radially polarized laser beam (not shown in Fig. 1a), an excitation that favours vertical electric dipoles exciting cylindrically diverging SPP waves propagating along the multilayer substrate surface[38-40,44].

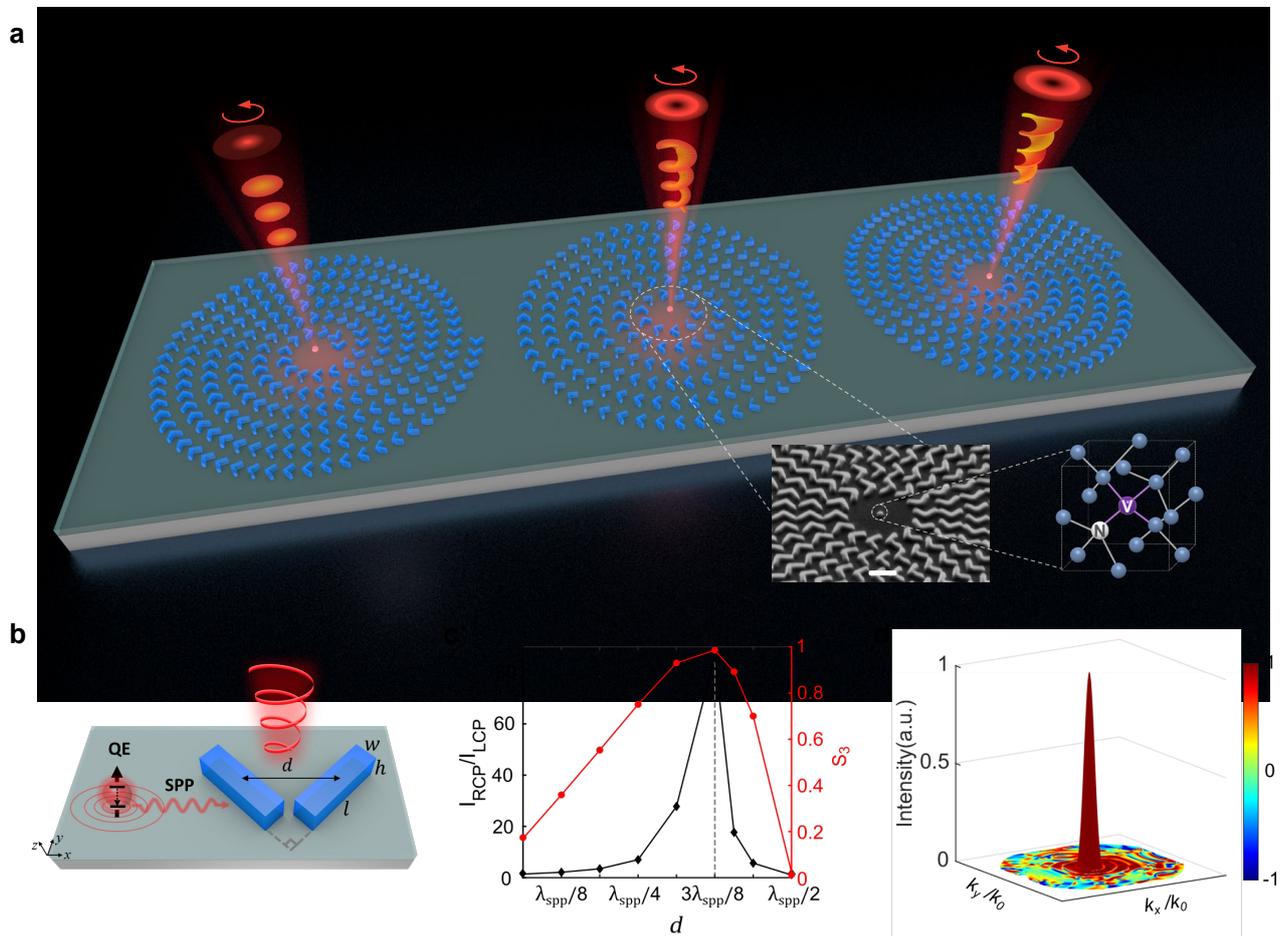

**Fig. 1 | Design of metasurfaces generating circularly polarized vortex beams. a**, Schematic of three QE-coupled quantum metasurfaces generating single-photon RCP single-mode vortex beams with different topological charges: $l_R = 0$, 1, and 2. The ND-NVs are surrounded by DONRs aligned along the OAM-corresponding trajectories. Insets: angular-view scanning electron microscopy (SEM) image of the central part of the fabricated metasurface along with the ND-NV schematic. Scale bar: 500 nm. **b**, Illustration of SAM

generation by DONR outcoupling of the QE-excited SPP wave. An individual DONR consists of two orthogonal nanorods with the width $w$ = 100 nm, length $l$ = 400 nm and height $h$ = 150 nm. **c**, The intensity ratio of RCP and LCP components (left axis) and Stokes parameter $S_3$ (right axis) as a function of the nanorod separation $d$ calculated at the radiation wavelength of 670 nm by considering the outcoupling of a plane SPP wave incident on a rectangular (7 × 6 $\mu$m$^2$) periodic array composed of DONRs. The array period is set equal to the SPP wavelength $\lambda_{spp} \cong 580$ nm to ensure the outcoupling in the normal (to the surface) direction. **d**, Three-dimensional (3D) representation of the superimposed far-field intensity and polarization distribution of the outcoupled beam produced by the rectangular DONR array with the optimized in (c) nanorod separation. The colour indicates the value of Stokes parameter $S_3$.

In general, the polarization of the photon emission outcoupled from QE-excited SPPs by a DONR can be represented as the SAM superposition state in the circularly polarized basis ($|R\rangle$ and $|L\rangle$)[45]:

$$|\Psi\rangle_{SAM} = C_1|R\rangle + C_2|L\rangle \ , \tag{1}$$

with $C_1$ being unequal to $C_2$ due to scattering anisotropy, unlike the case of outcoupling with nanoridges[40]. Arranging metasurface DONRs (oriented identically with respect to radially diverging SPP wave) along spiral trajectories separated by the SPP wavelength in the radial direction results in the corresponding OAM superposition states as follows[40]

$$|\Psi\rangle_{SUP} = C_1|R\rangle|l_R = m + 1\rangle + C_2|L\rangle|l_L = m - 1\rangle \ , \tag{2}$$

Here, $m$ is the number of spiral arms with $m$ < 0 for counterclockwise spirals. In our approach, we first design an individual DONR to obtain a high purity SAM state $|\Psi\rangle_{SAM}$ (e.g., with $C_1 \to 1$ and $C_2 \to 0$) and then arrange the DONR elements along the spiral trajectories to ensure the generation of a desirable OAM state superimposed with a given SAM state. For example, by outcoupling QE-excited SPPs with azimuthally arranged DONRs, each producing the RCP photons, along diverging (counterclockwise) spiral, concentric, and converging (clockwise) spiral trajectories, RCP vortex beams with correspondingly different OAMs, $l_R$ = 0, 1, and 2, are generated (Fig. 1a).

**DONR design.** The individual DONR represents a pair of orthogonal dielectric nanorods equally inclined (i.e., having ± 45°) with respect to the SPP propagation direction (Fig. 1b). In a qualitative treatment (Supplementary Note 1), the QE-excited SPP wave propagating along the *x*-axis is outcoupled by the DONR having the nanorod separation $d = \lambda_{spp}/4$ into the RCP(LCP) radiation produced by dipole moments generated along the long(short) sides of the nanorods (Supplementary Fig. 1a), with the former being the main contribution. The suppression of the unwanted LCP contribution can be very close to total for nanorods with large aspect (length-to-width, $l/w$) ratios. Thereby, the generation of high purity RCP photon emission with $C_1 \to 1$ and $C_2 \to 0$, is feasible, resulting in turn in the generation of the pure superposition state, having a desired OAM [equation (2)]. Likewise, a high purity LCP superposition state, i.e., with $C_1 \to 0$ and $C_2 \to 1$, can be generated with the DONR mirrored with respect to the *x*-axis (Supplementary Fig. 2a).

Nanofabrication of DONRs and the corresponding limitations on the nanorod dimensions, especially on the height-to-width ratio, $h/w$, result in relatively large nanorod dimensions as compared to the radiation wavelength. Consequently, the nanorods can no longer be treated within the electrostatic approximation and disregarding the inter-particle near-field coupling, as implicitly assumed in the qualitative treatment (Supplementary Note 1). Carefully conducted numerical simulations (Methods) confirmed that the suppression of the unwanted LCP contribution, $I_{RCP}/I_{LCP}$, can indeed be close to total, reaching 85, with the Stokes parameter $S_3$ attaining 0.98 for a slightly larger than expected (from the qualitative treatment) nanorod separation $d = 3\lambda_{spp}/8$ (Fig. 1c). The simulated performance of the plane SPP outcoupling by a rectangular DONR array with the optimized nanorod separation demonstrates an efficient SPP outcoupling into a well-collimated pure RCP photon beam (Fig. 1d). This excellent performance of an outcoupling DNAP array, efficiently performing the SPP-to-circular polarization transformation, has further been verified in the experiment with SPPs being excited by a periodic nanoridge grating illuminated with free-space laser light at 670 nm (Supplementary Note 2 and Fig. 3).

**SAM-OAM generation.** With the designed DONR at hand, it is straightforward to devise DONR location trajectories that would efficiently convert radially polarized and cylindrically diverging (QE-excited) SPP waves into circularly polarized photon emission, generating a desirable OAM state superimposed with the SAM state determined by the DONR orientation, i.e., generating a single-mode SAM-OAM encoded vortex beam [equation (2) and Supplementary Note 1]. The azimuthal arrangements with different spiral trajectories for the generation of three different SAM-OAM states of single photons are schematically shown in Fig. 1a and described in Table 1. The correspondingly designed configurations are displayed in Fig. 2a with the colour representing the SPP propagation phase variation (Supplementary Note 1).

**Table 1 | Three arrangements (M1, M2 and M3) for different SAM-OAM single-photon states generated by QE-coupled quantum metasurfaces**

| Metasurfaces arrangement | $|\Psi\rangle_{OAM}$ | $|\Psi\rangle_{SAM}$ |
|---|---|---|
| M1: Diverging spiral trajectory | $|\ell = 0\rangle$ | $|R\rangle$ |
| M2: Concentric trajectory | $|\ell = 1\rangle$ | $|R\rangle$ |
| M3: Converging spiral trajectory | $|\ell = 2\rangle$ | $|R\rangle$ |

Our numerical simulations (Methods) confirmed that the RCP states dominate the photon emission for all three configurations (Fig. 2b) as expected from the qualitative considerations above. The spatial overlaps of the RCP and LCP intensity distributions are found to be negligibly small: 0.3%, 1.7%, and 6.7%, respectively, - much smaller than those obtained previously with the QE-coupled metasurfaces consisting of continuous nanorods[38,40]. At the same time, the conservation of the total angular momentum noted previously[36,40] is manifested also with these DONR metasurfaces. For example, the diverging (counterclockwise) spiral results in the dominant RCP radiation concentrated in the far-field within a collimated bright spot with the uniform phase distribution, signifying the topological charge $\ell = 0$, while the LCP radiation (weaker by two orders of magnitude) is distributed in a doughnut shape with the phase distribution featuring a singularity with the topological charge $\ell = -2$ (Fig. 2b). Contrary to that, the converging (clockwise) spiral results in the strong RCP radiation shaped as a doughnut, carrying the topological charge $\ell = +2$, while the weak LCP radiation is concentrated within a phase-homogenous spot with the topological charge $\ell = 0$ (Fig. 2b).

The RCP relative contribution (weight) can be defined as the ratio of the emission into the RCP radiation, i.e., that with a sufficiently large Stokes parameter $S_3 > \xi$ ($\xi$ is the chiral threshold), to the total emitted power, and it can be estimated from the far-field intensity distribution as follows: $f = \frac{\sum_i I(S_3(i) > \xi)}{\sum_i I(i)}$. For the considered configurations, the RCP weight is rather high ($> 80\%$) even when requiring $S_3 > 0.9$, with the largest value of 98.2% found for the RCP generation with the diverging spiral trajectory while being somewhat lower for the concentric and converging spiral trajectories (Fig. 2c). The enhancement of RCP emission as compared with the QE emission without metasurfaces, when calculated as a function of the numerical aperture (NA) of the collecting radiation objective, emphasizes strong collimation of the RCP emission, exceeding three orders of magnitude for NA = 0.1 (Fig. 2d).

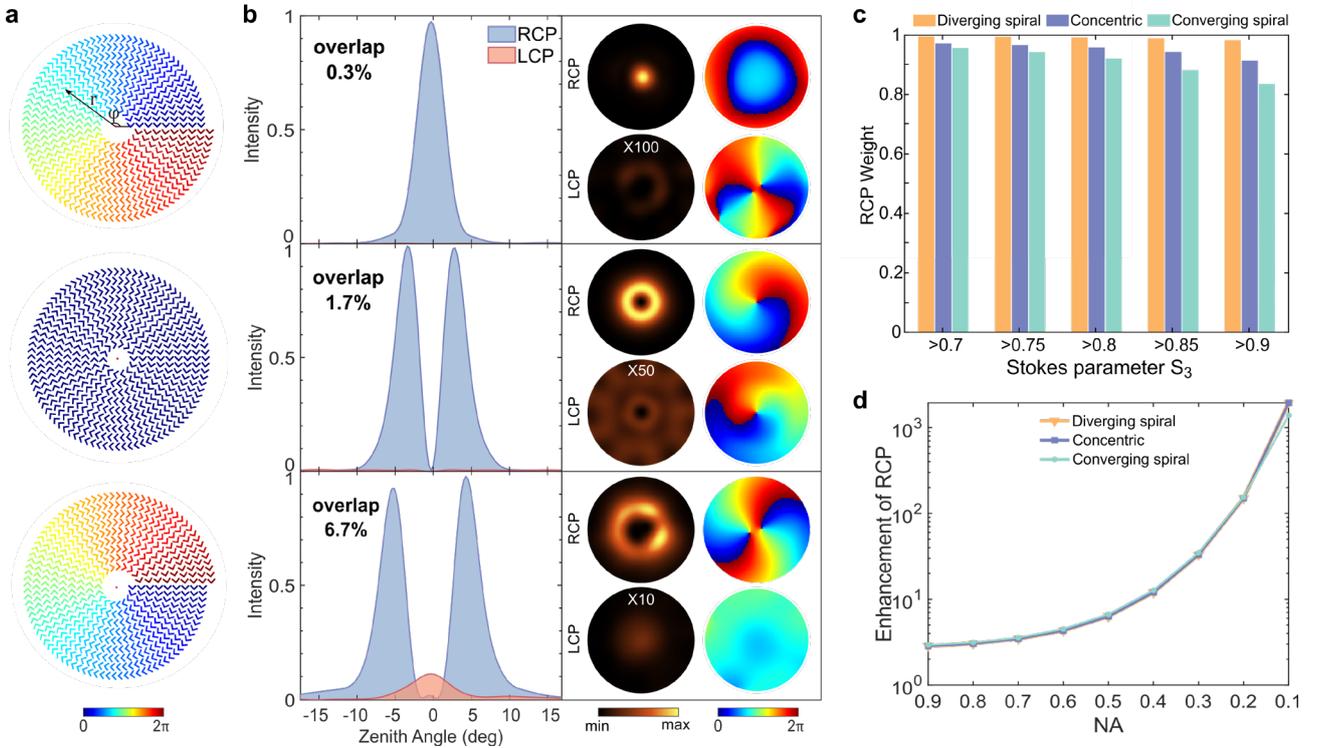

**Fig. 2 | Generation of circularly polarized single-mode vortex beams. a**, DONR azimuthal arrangements for generating high-purity circularly polarized vortices with (from top to bottom) diverging spiral, concentric, and converging spiral trajectories, which are coloured according to the SPP propagation phase. **b**, Simulated

intensity and phase distributions of the RCP and LCP contributions for the corresponding configurations. First column: the RCP and LCP intensity profiles with their spatial overlap quantified as 0.3, 1.7, and 6.7%, respectively. Second and third columns: the intensity (NA = 0.2) and corresponding phase distributions. The LCP intensity distributions have been enhanced by 100, 50, and 10 times, respectively to improve their visibility. **c**, The RCP weight indicating the relative RCP contribution in the total emission for the considered configurations and different chirality thresholds. **d**, Enhancement of the RCP emission with respect to the emission without metasurfaces as a function of the collecting objective NA.

We have verified the underlying physics of conversion from cylindrically diverging SPP waves into different SAM-OAM states with different DONR trajectories by utilizing an ideal DONR consisting of two orthogonal electric dipoles with the $\pi/2$-phase difference to compose the same spiral arrangements. All dipoles are considered to have the same strength with the phases being set in accordance with the SPP propagation phase. The simulation results demonstrated indeed the generation of perfect RCP (LCP contribution < 1%) vortex beams with the corresponding topological charges: $\ell = 0, 1,$ and $2$ (Supplementary Fig. 4). Similar results have been obtained with the mirrored DONR configuration producing perfect LCP vortices. This simple model, which does not require heavy calculations, might also be found useful when analysing the performance of other QE-coupled metasurfaces.

**Demonstration of single-mode SAM-OAM photon sources.** Following the design described above, various SAM-OAM photon sources were fabricated using NDs containing multiple NVs or single NV centres respectively (see Methods for details). The SEM images of the fabricated metasurfaces with DONRs lined up along different spiral trajectories evidence accurate positioning of NDs with multiple NVs (Fig. 3a) as well as high quality and homogeneity of the fabricated nanostructures (Supplementary Figs. 10 and 11). The performance of photon sources was characterized by evaluating the Stokes parameters ($S_0$, $S_1$, $S_2$, $S_3$), in which $S_0$ represents the total emission intensity, $S_1$ - the intensity difference between two orthogonal linear polarizations, $S_2$ - the intensity difference between other orthogonal linear polarizations, which are rotated by 45° with respect to the former one, and $S_3$ - the intensity difference between the right- or left-hand circular polarizations (all normalized to the total intensity obtained in each measurement)[38,46,47]. The Stokes parameters were directly measured with the photon emission passing through a quarter wave plate and linear polarizer (see Supplementary Note 3). We evaluate the chirality (i.e., the degree of circular polarization) of the observed photon emission using the chiral parameter $S_3$ weighted with the other parameters: $P_c = S_3/\sqrt{(S_1)^2 + (S_2)^2 + (S_3)^2}$. Far-field photon emission patterns feature well-collimated beams, with a very good correspondence between the simulated and experimentally observed intensity distribution (Fig. 3b and Supplementary Fig. 13). Note that the experimentally obtained divergence angles, 2.4º, 5º, and 7.1º, are very close to those expected from the simulations. The simulated and measured photon emission intensity and polarization distributions superimposed in 3D representations exhibit a very good correspondence (Fig. 3c), ascertaining the generation of single-mode SAM-OAM beams. To quantitatively characterize the performance with respect to the chirality of generated photon beams, we define the degree of circular polarization averaged over the intensity distribution: $\overline{P_c} = \sum_i P_c(i)I(i) / \sum_i I(i)$. The photon beam chirality can thus be evaluated as the average chirality $\overline{P_c}$ calculated from the high intensity data area (i.e., within the divergence angle indicated by the dashed line in Fig. 3b). This procedure results in the average beam chirality values of 0.934, 0.902, and 0.893 for the three realized (diverging spiral, concentric, and converging spiral) configurations, respectively. Moreover, the maximum chirality values are measured to be 0.996, 0.978, and 0.991 for these configurations, confirming further the generation of high-purity RCP beams (Fig. 3d).

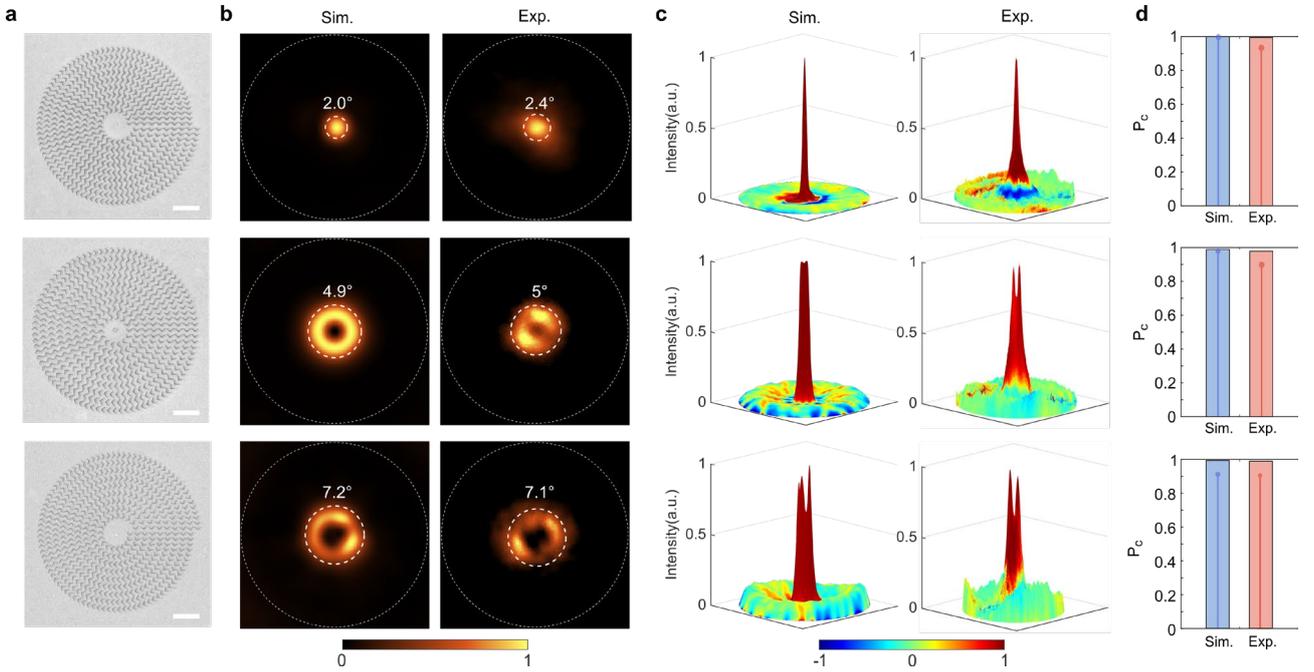

**Fig. 3 | Demonstration of single-mode SAM-OAM photon sources. a**, SEM images of the fabricated quantum metasurfaces for generating circularly polarized vortex photon beams. Scale bar: 2 $\mu$m for all SEM images. **b**, Simulated and experimentally measured intensity distributions of the photon emission from the corresponding quantum metasurfaces (NA=0.3). The white dashed circles encircle the areas corresponding to the full-width-at-half-maximum (FWHM) intensities. **c**, 3D representation of the superimposed intensity and polarization distributions. The height indicating the intensity and the colour indicates the degree of the circular polarization, with the red(blue) colour representing the RCP(LCP). **d**, The intensity-weighted $\overline{P_c}$ and the maximum $P_c$ in the high-intensity areas.

The OAM states of generated SAM-OAM photon beams were characterized by using a spatial light modulator (SLM), whose reflection was encoded with different orders of spiral phases, to demodulate phase profiles of OAM beams[40,45]. When changing the topological charge of the SLM phase distribution, the azimuthal intensity pattern of a photon beam reflected by the SLM should change in accordance with the topological charge of the generated vortex beam. A bright spot is expected to be formed when the topological charge of the SLM phase is opposite to that of the single-photon vortex beam (Supplementary Fig. 14). Our experimental characterization (Methods) demonstrated that the azimuthal intensity patterns observed for different SLM spiral phase distributions exhibit indeed well pronounced and concentrated bright spots for the corresponding topological charges of the SLM phase distribution (Supplementary Fig. 15), indicating thereby high quality of the fabricated SAM-OAM photon sources. We have also observed that the metasurfaces composed of the mirrored DONRs result in the generation of high-quality LCP photon Gaussian and vortex beams (Supplementary Fig. 16).

**Single-photon circularly polarized single-mode vortex beams.** Following the design, fabrication and characterization procedures described above (but utilizing NDs containing only single NV centres), we realized and characterized single-photon sources of well-collimated single-mode SAM-OAM beams. The main differences in the results of optical characterization are related to considerably weaker photon signals from the single-photon sources resulting in the optical images being more influenced by the background noise. Thus, the azimuthal intensity distributions observed for different SLM spiral phase distributions when characterizing three different single-photon sources exhibited essentially the same patterns (Fig. 4a), ascertaining also in this case, with the diverging, concentric, and converging spiral DONR trajectories, generation of the RCP vortex beams carrying the corresponding topological charges $\ell = 0, +1,$ and $+ 2$. The generated SAM-OAM photon modes were of high purity, especially the Gaussian RCP mode exhibited the impressive mode purity of 81%, with 94% being expected from simulations (Fig. 4b and 4c). The second-order correlation functions of photon emission measured before and after NV-ND coupling to the corresponding metasurface feature the zero-delay level $g^{(2)}(0) = 0.129$ and $0.141$, respectively (Fig. 4d), indicating excellent and stable single-photon properties. Also, the NV-ND lifetimes (Supplementary Fig. S18) and fluorescence spectra of photon emission (Fig. 4c) did not change significantly when measured before and after NV-ND coupling to the metasurface. The single-photon emission spectrum obtained after the NV-ND coupling features two zero-phonon lines (ZPLs), appearing at 575 and 637 nm and indicating, respectively, the charge state of NV dominated by negative ($NV^-$) and neutral ($NV^0$) charge state[47], and the emission peak at ~ 670 nm, which is the target wavelength of the designed metasurfaces.

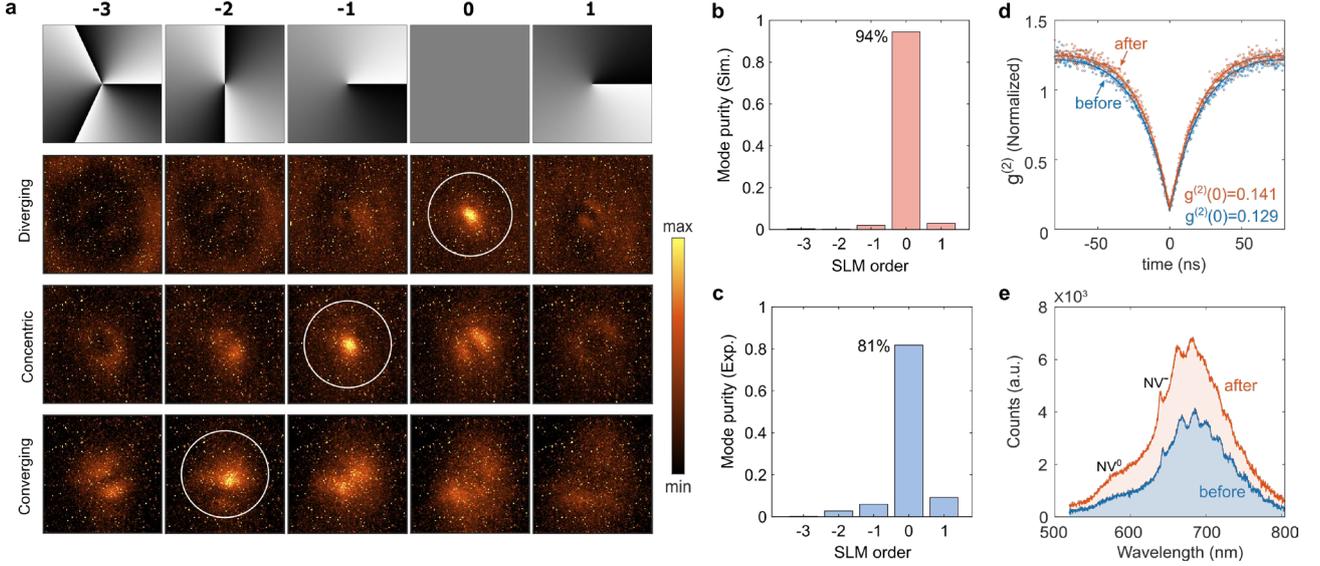

**Fig. 4 | Single-photon sources of circularly polarized single-mode vortex beams. a**, Intensity distributions of the single-photon emission obtained with different DONR spiral trajectories measured after the reflection from the SLM encoded with different orders of spiral phases. First row: phase distribution for the SLM holograms with the topological charges of -3, -2, -1, 0, and 1; second, third and fourth rows present the intensity distributions obtained with the diverging, concentric and converging spiral arrangements, respectively. **b**, **c**, Mode purity of the single-photon source simulated (**b**) and measured (**c**) with the diverging spiral trajectory. **d**, Second-order correlation function $g^{(2)}$ of single-photon emission measured before and after the ND-NV coupling with the metasurface using 100 μW of 532 nm continuous laser. **e**, Fluorescence spectrum of the single-photon emission measured before and after the ND-NV coupling.

## Discussion

In this work, we proposed the general approach for designing single-photon circularly polarized single-mode vortex beams, with the SAM and OAM states being separately controlled by the DONR configuration and array arrangement. The developed approach provides thereby a straightforward flexible way to precisely engineer the SAM-OAM states of single photons emitted by the DONR-based quantum metasurface. A set of room-temperature single-photon sources has been designed, fabricated, and characterized, demonstrating the generation of high-purity circularly polarized vortex beams by the fabricated QE-coupled quantum metasurfaces. Our successful experimental realizations of high-quality SAM-OAM encoded single-photon sources confirm the robustness of the developed approach with respect to small QE displacements with respect to the DONR arrangement as indicated by the numerical simulations (Supplementary Figs. 7 and 8). This approach can straightforwardly be extended to produce linearly polarized photon beams by interleaving two metasurfaces[48], i.e., DONR arrangements for providing the RCP and LCP photons. Furthermore, interleaving two metasurfaces for generating different SAM-OAM states opens a way for producing entangled SAM-OAM states, similarly to what has recently been demonstrated[40], but with a very significant difference: the total angular momentum of entangled states does not have to be the same – it can be arbitrarily different in accordance with the design of the individual metasurfaces (Supplementary Fig. 19). Finally, by combining the interleaving procedure[48] and recently established holographic formalism[36] the general approach developed in this work can further be extended to produce multiple, differently polarized, radiation channels, and enable thereby realization of high-dimensional quantum sources for advanced quantum photonic technologies. The on-chip room-temperature versatile handling of photon states opens thereby exciting perspectives for designing advanced single-photon sources and would contribute to further developments within chiral quantum optics, concerned with new functionalities such as spin/angular controlled photon switching and high-dimensional quantum information processing.

**Methods**

**Numerical simulations.** Numerical simulations of QE-coupled metasurfaces were performed with three-dimensional (3D) finite-difference time-domain (FDTD) method. An z-direction electric dipole was placed 30 nm atop the $SiO_2$ spacer (with the thickness of 20 nm) and silver film (with thickness of 150 nm) and in the centre of dielectric nanostructures (with height of 150 nm, refractive index 1.41). The wavelength of the dipole was set as 670 nm, fitting the emission peak wavelength of ND-NVs. The period of the metasurfaces was set be equals to the SPP wavelength ($\lambda_{spp} = 580$ nm with the effective SPP mode index $n_{spp} = 1.16$). The start radius $r_0$ is $1.7\lambda_{spp}$. The number of windings was $N = 9$, enabling efficiently outcoupling SPPs into well collimated far-field photon emission (Supplementary Figs. 5 and 6). A two-dimensional monitor 30 nm atop the configuration was employed to collect far-filed electric field by near- to far-field transformation method.

**Device fabrication.** Fabrication of QE-coupled metasurfaces followed a well-established sequence of technological steps (Supplementary Fig. 9) [38,46]. A 150 nm thick silver-film was first deposited on the silicon substrate by the thermal evaporation, followed by a 20-nm-thick $SiO_2$ layer being deposited by the magnetron sputtering method. Then a group of gold markers were fabricated on the substrate by the E-beam lithography (EBL), gold deposition, and the lift-off. Subsequently, ND-NVs were spin-coated with proper concentration. After that, the dark-field microscope image or the fluorescence scan image was taken to determine the position of, respectively, multiple or single ND-NVs with the help of alignment markers. It should be noted that the accurate positioning of DONR arrangements with respect to singles NV centres was more problematic as compared to the positioning of NDs with multiple NVs, because it had to be relied on fluorescence images (Supplementary Fig. 17) that were noisier than the dark-field images (Supplementary Fig. 9). After determining the position of ND-NVs with the precise alignment procedure, a negative photoresist (HSQ) was subsequently spin-coated at 4000 rpm, for 45s, and heated at the hotplate on 160 °C for 2 minutes to form a ~150 nm layer, which was examined by the AFM (atomic force microscope, NT-MDT NTEGRA). The EBL direct writing of metasurface patterns was then carried out on the resist around NV-NDs (JEOL-6490 system, accelerating voltage 30 kV). Finally, QE-coupled metasurfaces were obtained by development with the tetramethylammonium hydroxide solution (for 4 min) and isopropanol (60 s).

**Optical characterization.** Characterisation of QE-coupled metasurfaces was conducted using essentially the same experimental setup as in the previous experiments (Supplementary Fig. 12) [38,40].

*Excitation of single photons.* A linearly polarized 532 nm continuous wave (CW) or pulsed laser beam, after passing through a radial polarization converter (AR Coptix RPC), was used to driven NV-NDs to form a dominant dipole moment perpendicular to the substrate surface. The excitation and collection of emission were by the same objective with NA = 0.9 (The Olympus MPLFLN ×100).

*SAM and OAM states measurement.* SAM states: The Stokes parameters ($S_0$, $S_1$, $S_2$, $S_3$) were measured to characterize SAM states. $S_1$ and $S_2$ were measured by orienting the linear polarizer and taking Fourier plane images. $S_3$ was measured by the combination of a quarter-wave plate and a linear polarizer. OAM states: an SLM working at 670 nm was used to generate the computed holograms with different topological charges. As the SLM was only response to the linear polarized light, the RCP component generated from the QE-coupled metasurface was converted to the linearly polarized photons by a 45º-oriented quarter wave plate combined with a vertically placed linear polarizer. The reflected light from the incident light was filtered out by a set of dichroic mirrors (Semrock FF535-SDi01/FF552-Di02) and with long-pass filter 550 nm (Thorlabs FELH0550) and band-pass filter 676 nm (with 29 nm bandwidth, Edmund Optics).

*Single-photon characterization.* The photon emission was guided to spectrometer or single-photon detectors for the measurement of correlation, spectrum, and fluorescence image via the flip mirror. Fluorescence imaging (for ND-NV positioning process) were performed by a synchronous movement of piezo-stage with mounted sample and projecting the collected fluorescence emission to the avalanche photo diodes (APDs). Decay-rate measurements was carried out with a pulsed laser (Pico Quant LDH-PFA-530L) at 1 MHz together with APD1 (τ -SPAD, PicoQuant). Fluorescence spectra were measured using the spectrometer (Andor Ultra 888 USB3-BV) operating within 540-820 nm. Second-order correlation was recorded by registering the temporal delay between photon detection events between APD1 and APD2 in a start-stop configuration, using an electronic timing box (Picharp-300; Pico quant).

**Acknowledgements**

The authors acknowledge the support from National Natural Science Foundation of China (Grant No. 62105150 and No. 52120105009), European Union's Horizon Europe research and innovation programme under the Marie Skłodowska-Curie Action (Grant agreement No. 101064471), Natural Science Foundation of Jiangsu Province (BK20210289), Shanghai Key Fundamental Research Grant (No. 20JC1414800), State Key Laboratory of Advanced Optical Communication Systems Networks of China (2022GZKF023), and Villum Kann Rasmussen Foundation (Award in Technical and Natural Sciences 2019).

**Author contributions**

S.I.B. conceived the idea. X.J.L. and Y.H.K. performed theoretical modelling. X.J.L. with assistance from Y.H.K. and D.K. fabricated samples. X.J.L. and Y.H.K. with assistance from S.K. and D.K. performed experimental measurement. X.J.L., Y.H.K., S. K., and S.I.B. analysed the data. S.I.B., Y.H.K., and C.Y.Z supervised the project. X.J.L., Y.H.K. and S.I.B wrote the manuscript with contributions from all authors.

**Competing interests**

The authors declare no competing interests.

# Supplementary information

# Single-photon circularly polarized single-photon vortex beams

Xujing Liu[1,2], Yinhui Kan[2], Shailesh Kumar[2], Danylo Komisar[2], Changying Zhao[1], Sergey I. Bozhevolnyi[2],

[1]Institute of Engineering Thermophysics,

Shanghai Jiao Tong University, Shanghai, 200240, China

[2]Centre for Nano Optics,

University of Southern Denmark, DK-5230 Odense M, Denmark

**Supplementary Note 1 | Theoretical analysis of SPP scattering into circularly polarized light**

(1) Producing RCP with anisotropic nanobricks

We first introduce conventions for describing the light fields, a plane electromagnetic wave is of the form[1]:

$$\boldsymbol{E}_{pr} \sim \boldsymbol{E}_0 \exp\{i(\omega t - \boldsymbol{k} \cdot \boldsymbol{r})\} \rightarrow \boldsymbol{E}_{pr} \sim \boldsymbol{E}_0 \exp(-i\boldsymbol{k} \cdot \boldsymbol{r}) \quad (S1)$$

where $\boldsymbol{E}_0$ is the electric field vector, $\omega$ is the angular frequency, the wave vector is, $\boldsymbol{k} = 2\pi/\lambda$, $\lambda$ is the wavelength of the wave. For a right-hand circularly polarized (RCP) wave travelling in the +z direction:

$$\boldsymbol{E}_{pr} \sim E_0 \frac{1}{\sqrt{2}} \begin{pmatrix} 1 \\ -i \end{pmatrix} \exp\{i(\omega t - \boldsymbol{k} \cdot \boldsymbol{r})\} \sim E_0 \begin{pmatrix} 1 \\ -i \end{pmatrix} \exp(-ikz) \quad (S2)$$

We consider the SPP incidence on a doublet of (extremely) anisotropic scatterers displaced by a quarter of the SPP wavelength along the propagation direction as shown in Fig. S1(a), considering only the in-plane (x, y) field components, the incident SPP field has only one component:

$$\boldsymbol{E}_{SPP} \sim E_0(1, 0) \exp(-ik_{SPP}x) \quad (S3)$$

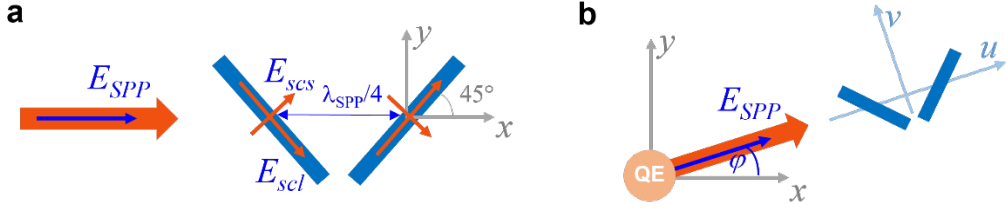

**Supplementary Figure 1 | Schematic illustration of generating RCP. a**, Normal incident SPP scattered by a doublet of anisotropic scatterers. **b**, Regime of QE-excited SPPs propagating under an azimuthal angle $\varphi$.

The (largest) field component scattered by this doublet in the z direction can be written as follows:

$$\boldsymbol{E}_{scl} \sim C_1 \begin{pmatrix} 1 + \exp\left(-i\frac{\pi}{2}\right) \\ -1 + \exp\left(-i\frac{\pi}{2}\right) \end{pmatrix} = C_1 \begin{pmatrix} 1-i \\ -1-i \end{pmatrix} = C_1(1-i) \cdot \begin{pmatrix} 1 \\ -i \end{pmatrix} \sim C_1 \cdot e^{-i\frac{\pi}{4}} \cdot \begin{pmatrix} 1 \\ -i \end{pmatrix} \quad (S4)$$

If we consider the contributions from the short side of a nanobrick, the scattered field component is relatively small and can then be written as follows:

$$\boldsymbol{E}_{scs} \sim C_2 \begin{pmatrix} 1 + \exp\left(-i\frac{\pi}{2}\right) \\ 1 - \exp\left(-i\frac{\pi}{2}\right) \end{pmatrix} = C_2 \begin{pmatrix} 1-i \\ 1+i \end{pmatrix} \sim C_2 \cdot e^{-i\frac{\pi}{4}} \cdot \begin{pmatrix} 1 \\ i \end{pmatrix} \quad (S5)$$

Conclusions:

1) The largest scattered field components produce an RCP wave travelling in the $+z$ direction (i.e., towards us), while the minor scattered field components result in an LCP wave travelling in the $+z$ direction ($C_1 \gg C_2$).

2) The net scattering in both cases is coming from the doublet centre as reflected by the factor $e^{-i\frac{\pi}{4}}$.

(2) <u>Producing RCP vortices</u>

Considering the SPP field propagating under an azimuthal angle $\varphi$ with the $x$-axis [Fig. S1(b)], the largest contribution to the scattered field can be expressed as follows:

$$\boldsymbol{E}_{scl}(\varphi) \sim C_1 \cdot \begin{bmatrix} \cos\varphi & -\sin\varphi \\ \sin\varphi & \cos\varphi \end{bmatrix} \cdot \begin{pmatrix} 1 \\ -i \end{pmatrix} = C_1 \begin{pmatrix} \cos\varphi + i\sin\varphi \\ \sin\varphi - i\cos\varphi \end{pmatrix} = C_1 e^{i\varphi} \begin{pmatrix} 1 \\ -i \end{pmatrix} \tag{S6}$$

Consequently, the minor contribution to the scattered field can be expressed as follows:

$$\boldsymbol{E}_{scs}(\varphi) \sim C_2 \cdot \begin{bmatrix} \cos\varphi & -\sin\varphi \\ \sin\varphi & \cos\varphi \end{bmatrix} \cdot \begin{pmatrix} 1 \\ i \end{pmatrix} = C_2 \begin{pmatrix} \cos\varphi - i\sin\varphi \\ \sin\varphi + i\cos\varphi \end{pmatrix} = C_2 e^{-i\varphi} \begin{pmatrix} 1 \\ i \end{pmatrix} \tag{S7}$$

Conclusions:

1) The largest scattered field components originating from concentric circles of doublets (separated by the SPP wavelength) produce an RCP vortex field with the $+1$ topological charge travelling in the $+z$ direction (i.e., towards us), while the minor scattered field components result in an LCP vortex field with the $-1$ topological charge.

2) Considering the largest scattered field components producing RCP fields, in order to produce an RCP wave (without topological charges) one should have a diverging (from the centre) spiral of nano-bricks with the distance from the centre (QE) increasing with the angle $\varphi$:

$$\varphi = k_{SPP} \cdot \delta r(\varphi) \rightarrow r(\varphi) = r_0 + \frac{\varphi}{k_{SPP}} \tag{S8}$$

In order to add the compensating phase:

$$\boldsymbol{E}_{SPP}(\varphi) \sim E_0 \begin{pmatrix} \cos\varphi \\ \sin\varphi \end{pmatrix} \exp(-ik_{SPP}r) \sim \exp(-i\varphi) \tag{S9}$$

3) Using converging (towards the centre) spiral would result in an RCP vortex field with the $+2$ topological charge travelling in the $+z$ direction.

4) In order to produce an RCP vortex with the $-1$ topological charge, one should use a twice fast diverging spiral:

$$2\varphi = k_{SPP} \cdot \delta r(\varphi) \rightarrow r(\varphi) = r_0 + \frac{2\varphi}{k_{SPP}} \tag{S10}$$

The compensating phase is:

$$\boldsymbol{E}_{SPP}(\varphi) \sim E_0 \begin{pmatrix} \cos\varphi \\ \sin\varphi \end{pmatrix} \exp(-ik_{SPP}r) \sim \exp(-i2\varphi) \tag{S11}$$

(3) <u>Producing LCP with anisotropic nanobricks</u>

We consider then the SPP incidence on a mirrored doublet of (extremely) anisotropic scatterers displaced by a quarter of the SPP wavelength along the propagation direction as shown in Fig. S2(a):

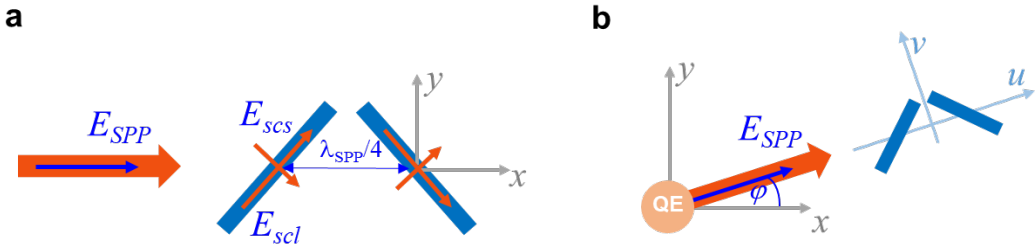

**Supplementary Figure 2 | Schematic illustration of generating LCP. a**, Normal incident SPP scattered by a doublet of anisotropic scatterers. **b**, Regime of QE-excited SPPs propagating under an azimuthal angle $\varphi$.

The (largest) field scattered by this doublet in the z direction can be written as follows:

$$\boldsymbol{E}_{scl} \sim C_1 \left(1 + \exp\left(-i\frac{\pi}{2}\right), -1 + \exp\left(-i\frac{\pi}{2}\right)\right) = C_1 \begin{pmatrix} 1-i \\ -1-i \end{pmatrix}$$

$$= C_1(1-i) \cdot \begin{pmatrix} \frac{1}{-1-i} \\ \frac{1-i}{1-i} \end{pmatrix} = C_1(1-i) \cdot \begin{pmatrix} 1 \\ -i \end{pmatrix} \sim C_1 \cdot e^{-i\frac{\pi}{4}} \cdot \begin{pmatrix} 1 \\ -i \end{pmatrix} \tag{S12}$$

If we consider the contributions from the short side of a nanobrick, the scattered field can then be written as follows:

$$\boldsymbol{E}_{scs} \sim C_2 \left(1 + \exp\left(-i\frac{\pi}{2}\right), -1 + \exp\left(-i\frac{\pi}{2}\right)\right) = C_2 \begin{pmatrix} 1-i \\ -1-i \end{pmatrix} \sim C_2 \cdot e^{-i\frac{\pi}{4}} \cdot \begin{pmatrix} 1 \\ -i \end{pmatrix} \tag{S13}$$

(4) <u>Producing LCP vortices</u>

Considering the SPP field propagating under the angle $\varphi$ with the *x*-axis as shown in Fig. S2(b), the largest contribution to the scattered field can be expressed as follows:

$$\boldsymbol{E}_{scl}(\varphi) \sim C_1 \cdot \begin{bmatrix} \cos\varphi & -\sin\varphi \\ \sin\varphi & \cos\varphi \end{bmatrix} \cdot \begin{pmatrix} 1 \\ i \end{pmatrix} = C_1 \begin{pmatrix} \cos\varphi - i\sin\varphi \\ \sin\varphi + i\cos\varphi \end{pmatrix} = C_1 e^{-i\varphi} \begin{pmatrix} 1 \\ i \end{pmatrix} \tag{S14}$$

Consequently, the minor contribution to the scattered field can be expressed as follows:

$$\boldsymbol{E}_{scs}(\varphi) \sim C_2 \cdot \begin{bmatrix} \cos\varphi & -\sin\varphi \\ \sin\varphi & \cos\varphi \end{bmatrix} \cdot \begin{pmatrix} 1 \\ -i \end{pmatrix} = C_2 \begin{pmatrix} \cos\varphi + i\sin\varphi \\ \sin\varphi - i\cos\varphi \end{pmatrix} = C_2 e^{i\varphi} \begin{pmatrix} 1 \\ -i \end{pmatrix} \tag{S15}$$

Conclusions:

1) The largest scattered field components originating from concentric circles of doublets (separated by the SPP wavelength) produce an LCP vortex field with the $-1$ topological charge travelling in the $+z$ direction, while the minor scattered field components result in an RCP vortex field with the $+1$ topological charge.

2) Considering the largest scattered field components producing LCP fields, in order to produce an LCP wave (without topological charges) one should have a converging (to the centre) spiral of nano-bricks with the distance from the centre (QE) decreasing with the angle $\varphi$:

$$\varphi = -k_{SPP} \cdot \delta r(\varphi) \rightarrow r(\varphi) = r_0 - \frac{\varphi}{k_{SPP}} \tag{S16}$$

In order to add the compensating phase:

$$\boldsymbol{E}_{SPP}(\varphi) \sim E_0(\cos\varphi, \sin\varphi)\exp(-ik_{SPP}r) \sim \exp(i\varphi) \tag{S17}$$

3) Using diverging (from the centre) spiral would result in an LCP vortex field with the $-2$ topological charge travelling in the $+z$ direction (with the strategy for producing positive topological charges being reversed).

(5) Concluding remarks

The ratio between the strongest and weakest scattered field components is determined by the length-to-width ratio of nanorods (nanobricks or anisotropic scatterers) but is not directly proportional to that. The corresponding dependence is influenced by the nanobrick shape and includes the nanobrick height and the refractive index as well. Nanofabrication along with the design principle limits the largest length-to-width ratio that can realistically be fabricated. For example, it is often very difficult to fabricate a nanobrick with the width smaller than the height. Given the nanobrick separation equal to a quarter of the SPP wavelength implies that the nanobrick length is limited to $\sqrt{2}\lambda_{SPP}/4$ to avoid the overlap of two orthogonal nanobricks (Fig. S1). Finally, the scattering strength should not be too small for the out-of-plane SPP scattering being stronger than the SPP absorption, and that requires to fabricate nanobricks with sufficiently large volumes, i.e., with sufficiently large heights. It is then clear that the overall optimization would require to use nanobrick materials with larger refractive indexes, relaxing thereby the requirements of large volumes (the scattering strength increases with the refractive index contrast). Our choice of the nanobrick parameters [Fig. 1(b) in the main text] is a result of numerical optimization involving extensive simulations and backed up with the control experiments.

**Supplementary Note 2 | Experiment of SPP scattering into circularly polarized light**

With the numerical simulations showing that the surface plasmon polaritons could be efficiently outcoupled to free-space circular polarized light by the designed anisotropic element, here, we conducted experimental characterization to investigate the performance of anisotropic units, which is the elementary component for the SPP coupler (metasurface B shown in Fig. S3(a)). The excitation of propagating SPPs is realized by a grating A (HSQ) on $SiO_2$ (20 nm) /Ag (150 nm) substrate. The period is 580 nm with the width of 140 nm, enabling the excitation of SPP with 670 nm laser light. The distance between grating A and metasurface B is 10 μm, considering the trade-off between the overlap of incident beam and outcoupling light and the SPP loss along propagation. A wavelength-tuneable continuous-wave laser (NKT, SuperK EXTREME/FIANIUM) is used as the incident light after converting to TM wave with a polarizer, then focused on the centre of the grating A. As clearly shown in Fig. S3(b), SPP can be excited and then coupling out by metasurface B, confirming its ability for acting as an efficient SPP outcoupler. To characterize the polarized states, a quarter wave plate and linear polarizer is inserted before the CMOS camera to measure the Stokes parameter $S_3$ (shown in the Fig. S3(c)). Fig. S3(d) shows the 2D distribution of $S_3$. Reflected light of grating A keeps the linear polarization, while the outcoupling emission from metasurface B is circular polarized with $S_3 = 0.98$. It proves that the SPPs can be efficiently outcoupled to free-space circular polarized light by the designed anisotropic element.

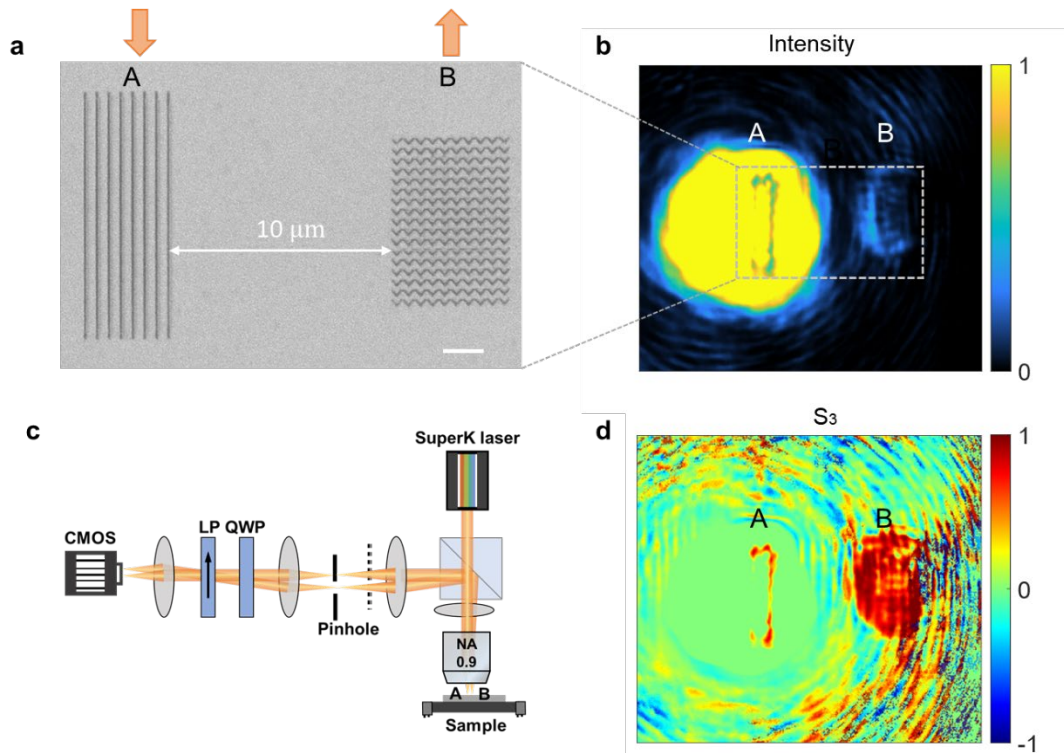

**Supplementary Figure 3 | Experimental demonstration of RCP generation with optimized element. a**, SEM images of the fabricated grating-metasurface. Scale bar: 2 μm. **b**, The real plane under TM incidence at 670 nm. **c**, Corresponding experimental setup. **d**, Measured Stokes parameter $S_3$.

## Supplementary Note 3 | Polarization measurement

The Stokes polarization parameters are directly measurable based on intensity quantities. The emitted light pass through a quarter wave plate (retardation angle $\varphi$), then followed by a linear polarizer with its transmission axis aligned at an angle $\theta$ to the $x$ axis. The emitted light of the single photon source can be decomposed as:

$$E_x(t) = E_{0x}e^{i\delta_x} \cdot e^{i\omega t} = E_x e^{i\omega t} \tag{S18}$$

$$E_y(t) = E_{0y}e^{i\delta_y} \cdot e^{i\omega t} E_y e^{i\omega t} \tag{S19}$$

where $E_x$ and $E_y$ are the complex amplitude. The Stokes parameter for a plane wave can be obtained from:

$$S_0 = E_x E_x^* + E_y E_y^*$$

$$S_1 = E_x E_x^* - E_y E_y^*$$

$$S_2 = E_x E_y^* + E_y E_x^*$$

$$S_3 = i(E_x E_x^* + E_y E_y^*) \tag{S20}$$

where $E_x^*$ and $E_y^*$ are the complex conjugates of $E_x$ and $E_y$.

The intensity $I(\varphi, \theta)$ of the emitted light is a function of retardation angle $\varphi$ and polarizer alignment angle $\theta$:

$$I(\varphi, \theta) = E_x E_x^* \cos^2\theta + E_y E_y^* \sin^2\theta + E_x^* E_y e^{-i\varphi}\sin\theta\cos\theta + E_x E_y^* e^{-i\varphi}\sin\theta\cos\theta \tag{S21}$$

The intensities at four different pairs of $\varphi$ and $\theta$ are measured to calculate the four Stokes parameter. The first three Stokes parameter is measured by rotating the polarizer to angle $\theta = 0°, 90°, 45°, \text{and} -45°$ respectively (remove the quarter wave plate). The final parameter $S_3$ is measured by the quarter wave retarder ($\theta = \pm 45°, \varphi = 90°$) and linear polarizer ($\theta = 0°$). The Stokes parameter is derived as:

$$S_0 = I(0°, 0°) + I(90°, 0°)$$

$$S_1 = I(0°, 0°) - I(90°, 0°)$$

$$S_2 = I(45°, 0°) - I(-45°, 0°)$$

$$S_3 = I(45°, 90°) - I(-45°, 90°) \tag{S22}$$

The degree of circular polarization of photon emission is defined as

$$P_c = S_3/\sqrt{(S_1)^2 + (S_2)^2 + (S_3)^2} \tag{S23}$$

in which $S_1$, $S_2$, and $S_3$ are the Stokes parameters normalized to the corresponding total intensity ($S_0$) obtained in each measurement.

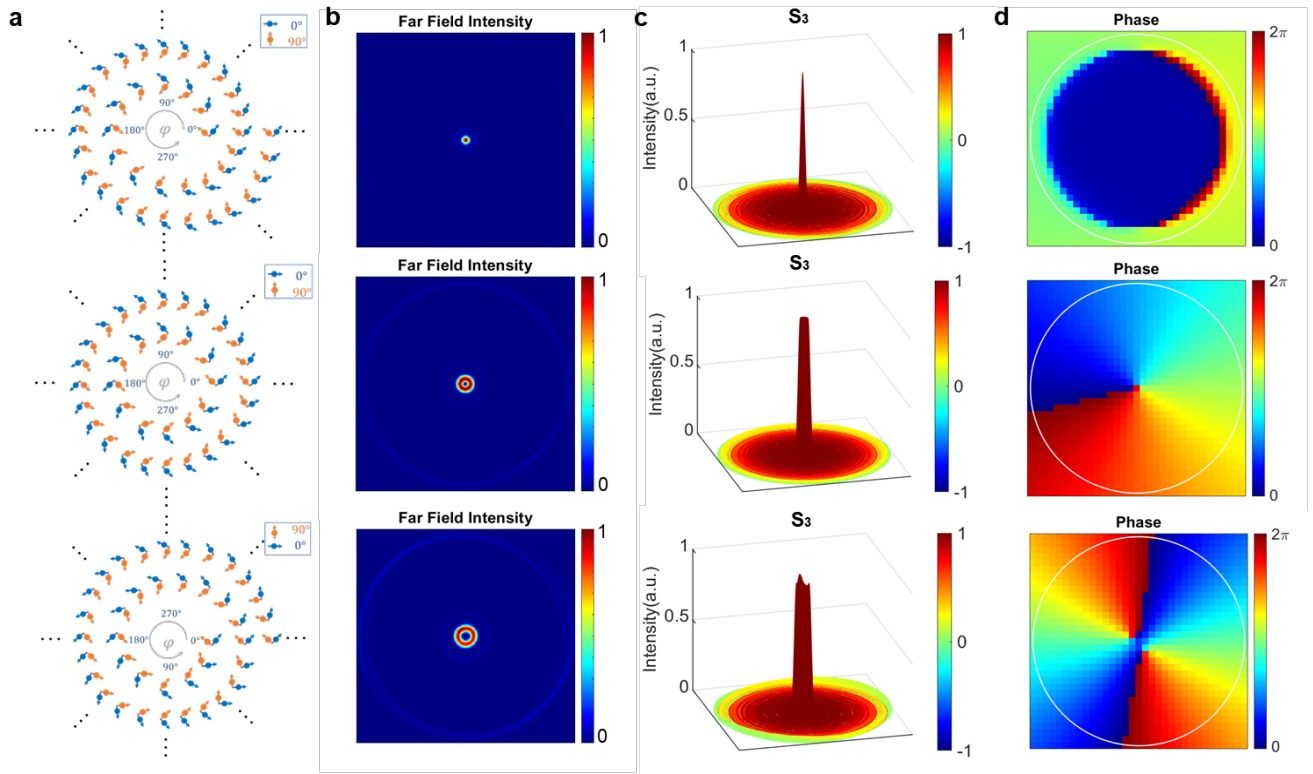

**Supplementary Figure 4 | Simulated results with dipole arrays, mimicking the designed configuration. a,** Schematic of the orthogonal dipole unit arranged with diverging spiral, concentric circle and converging spiral trajectories. There is a phase shift between the orthogonal dipole for each unit, forming perfect circular polarization. **b,** Far-field angular intensity distribution (in the Fourier plane). **c,** 3D superimposed intensity and polarization states, the colour shows the Stokes parameter $S_3$. **d,** Phase distribution of the decomposed RCP component. The white circle shows the collection angle of NA = 0.1.

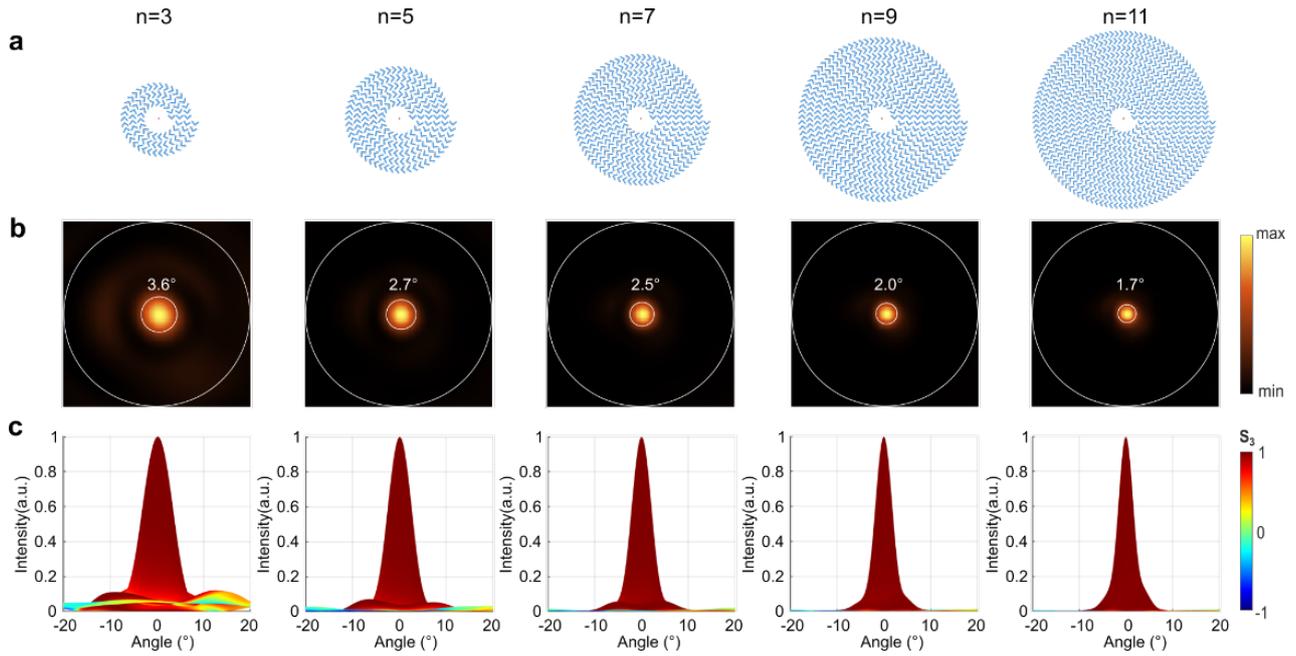

**Supplementary Figure 5 | Influence of the number of rings on the performance. a**, Top view of diverging Archimedean spiral configurations with different number windings (n = 3, 5, 7, 9, and 11). **b**, Far-field intensity distribution with numerical aperture NA = 0.2. The inner circle denotes the divergence angle. **c**, The superimposed intensity and degree of circular polarization (the colour represents Stokes parameter $S_3$).

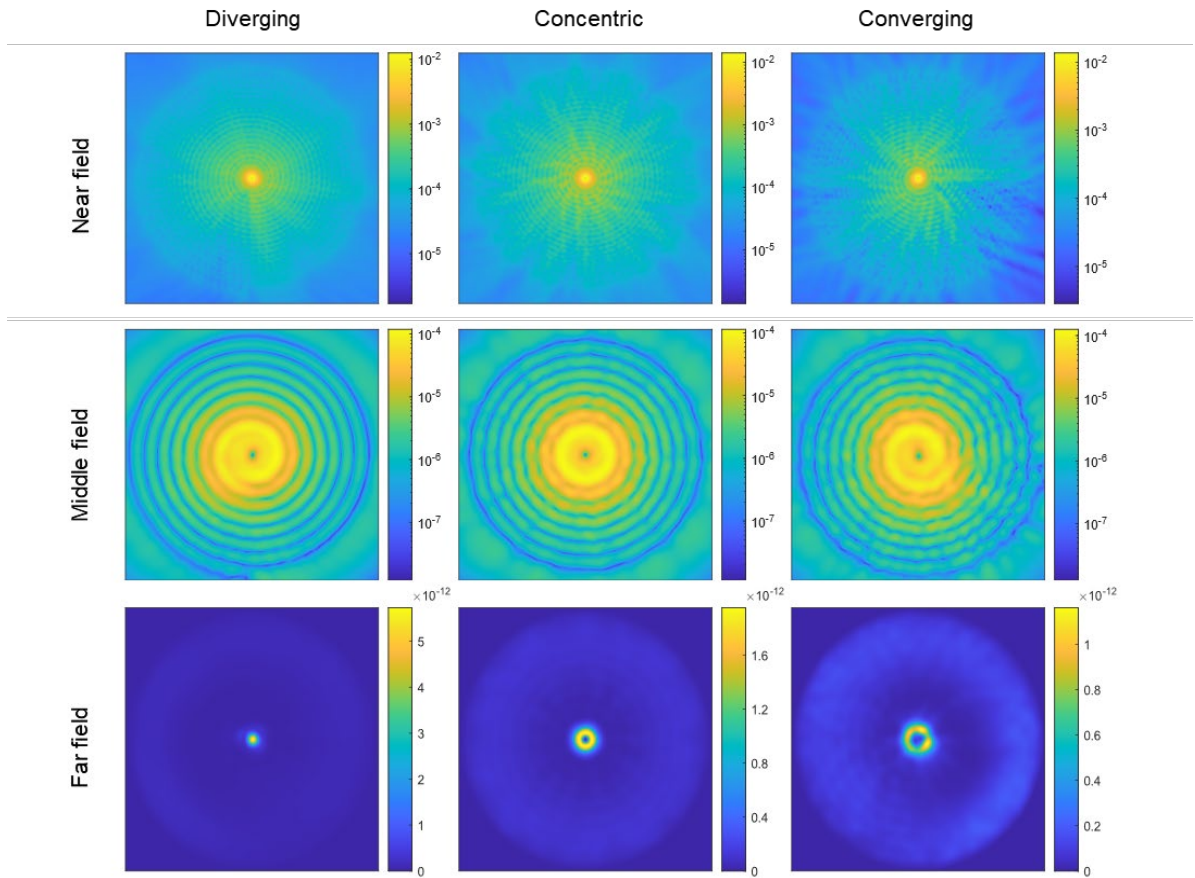

**Supplementary Figure 6 | Numerical simulations of energy flow (Poynting vector $P_z$) with designed diverging, concentric, and converging configurations.** In the near field, monitor is 30 nm away from metasurface. In the middle field, monitor is 900 nm away from metasurface, the far field is calculated with near to far field transformation. The results present the evolution of energy flow from the dipole source: In the near field, the photon emission is dominated by the dipole source itself; In the middle field, it shows the interactions between the dipole-excited SPPs with the surrounding nanostructures; In the far field, the dipole-excited SPPs are coupled into free space photon emission with different phase profiles (topological charges), showing a bright spot or doughnuts.

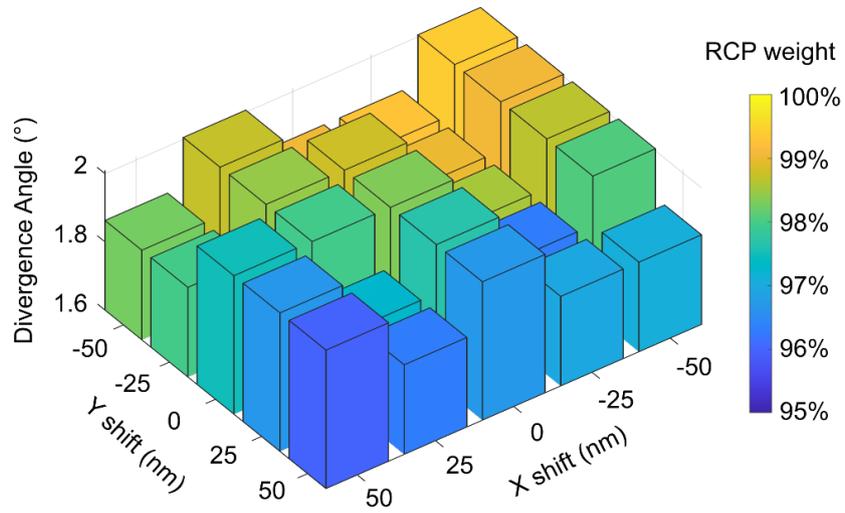

**Supplementary Figure 7 | Influence of the position shift of QEs on the performance with diverging spiral configuration.** The circular polarization purity maintains high value (lager than 95%) and the divergence angle varies from 1.86° to 2°, which are all robust to the position shift of QEs within 50 nm.

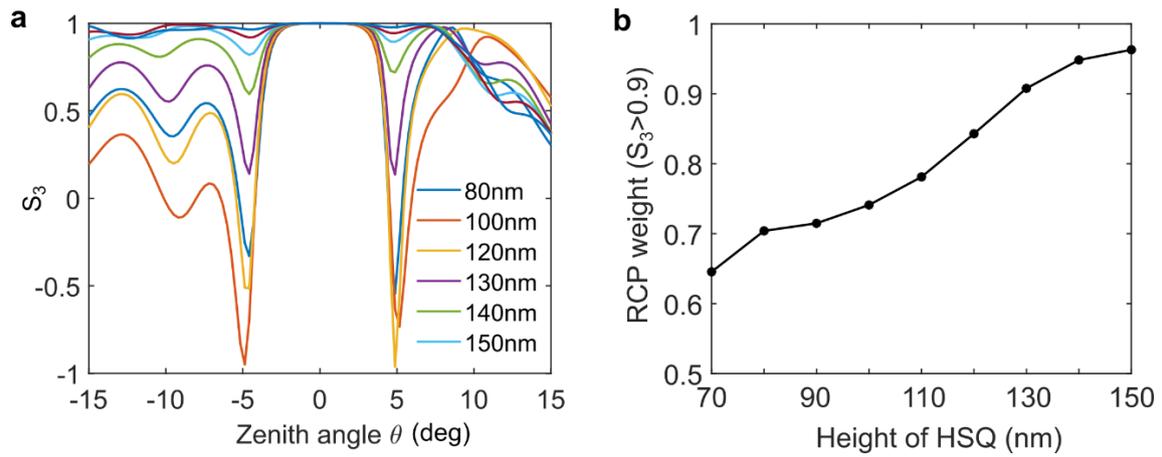

**Supplementary Figure 8 | Influence of the height of HSQ on the performance of circular polarized states.**
**a**, Simulated Stokes parameter $S_3$ within zenith angle $\pm15°$. **b**, Variation of RCP purity with the height of HSQ.

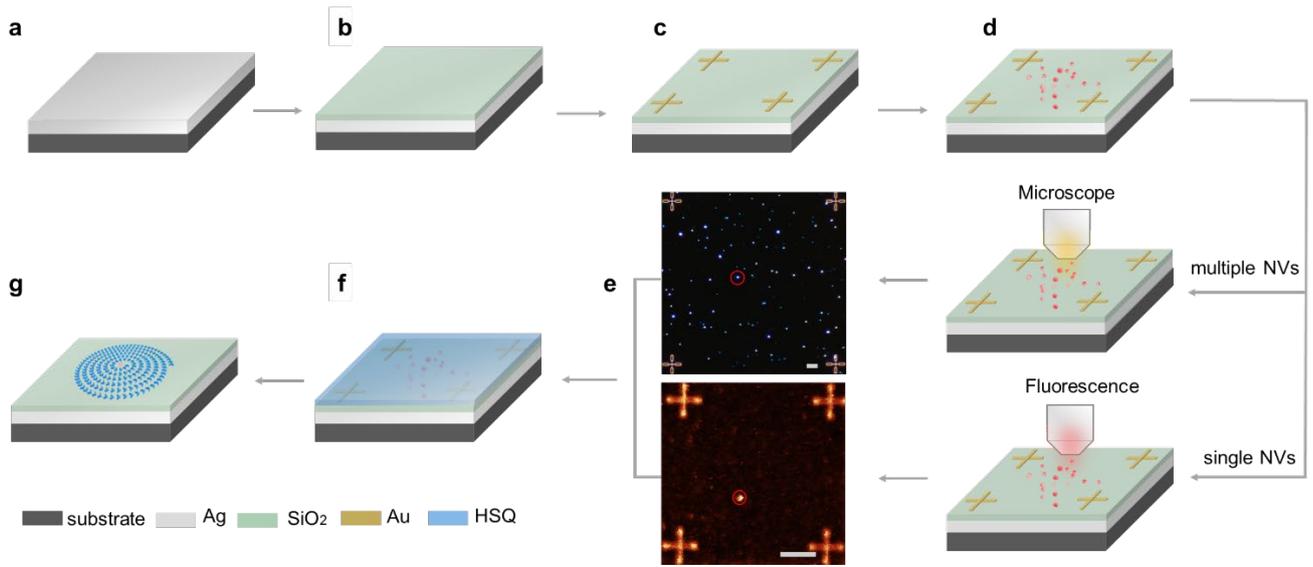

**Supplementary Figure 9 | Fabrication process of photon sources with multiple ND-NVs and single ND-NV. a**, Deposition of 150 nm Ag on the silicon substrate. **b**, 20 nm SiO$_2$ deposition. **c**, The alignment gold markers are fabricated by using EBL, gold deposition, and lift-off process. **d**, Spin coating nanodiamond containing NVs. **e**, Determine the position of ND-NVs. A dark-field microscope image is taken to determine the position of ND-NVs. Single-photon ND-NVs is searched by the fluorescence scan with a radially polarized excitation laser beam (532 nm), the position of which is determined by the fluorescence image with the help of alignment markers at the corners of a 27×27 μm$^2$ area. Scale bar: 5 μm. **f**, Spin coating HSQ and baked at the hotplate to form 150 nm HSQ layer. **g**, Metasurfaces are fabricated around the ND-NVs by EBL and the precise alignment procedure. The alignment and position determining method can be found in Ref [2].

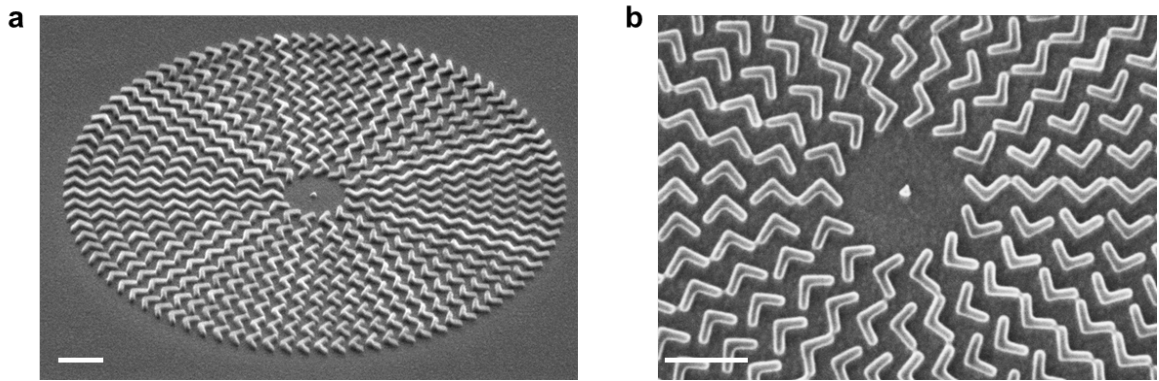

**Supplementary Figure 10 | SEM of the fabricated devices. a**, Angular-view scanning electron micrograph. **b**, Top-view SEM images of the centre part. ND-NVs is positioned in the centre of matasurface. Scale bar: 1 µm.

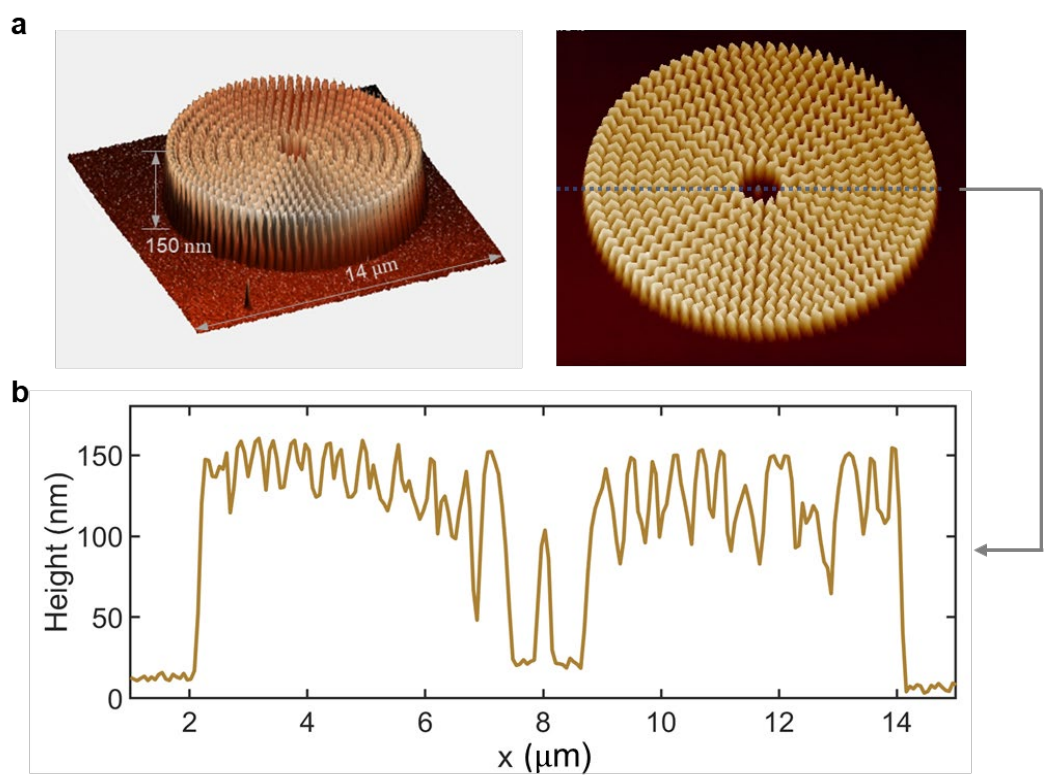

**Supplementary Figure 11 | AFM of the fabricated devices. a**, 3D AFM schematic of the fabricated sample. **b**, Height of HSQ in the cross section.

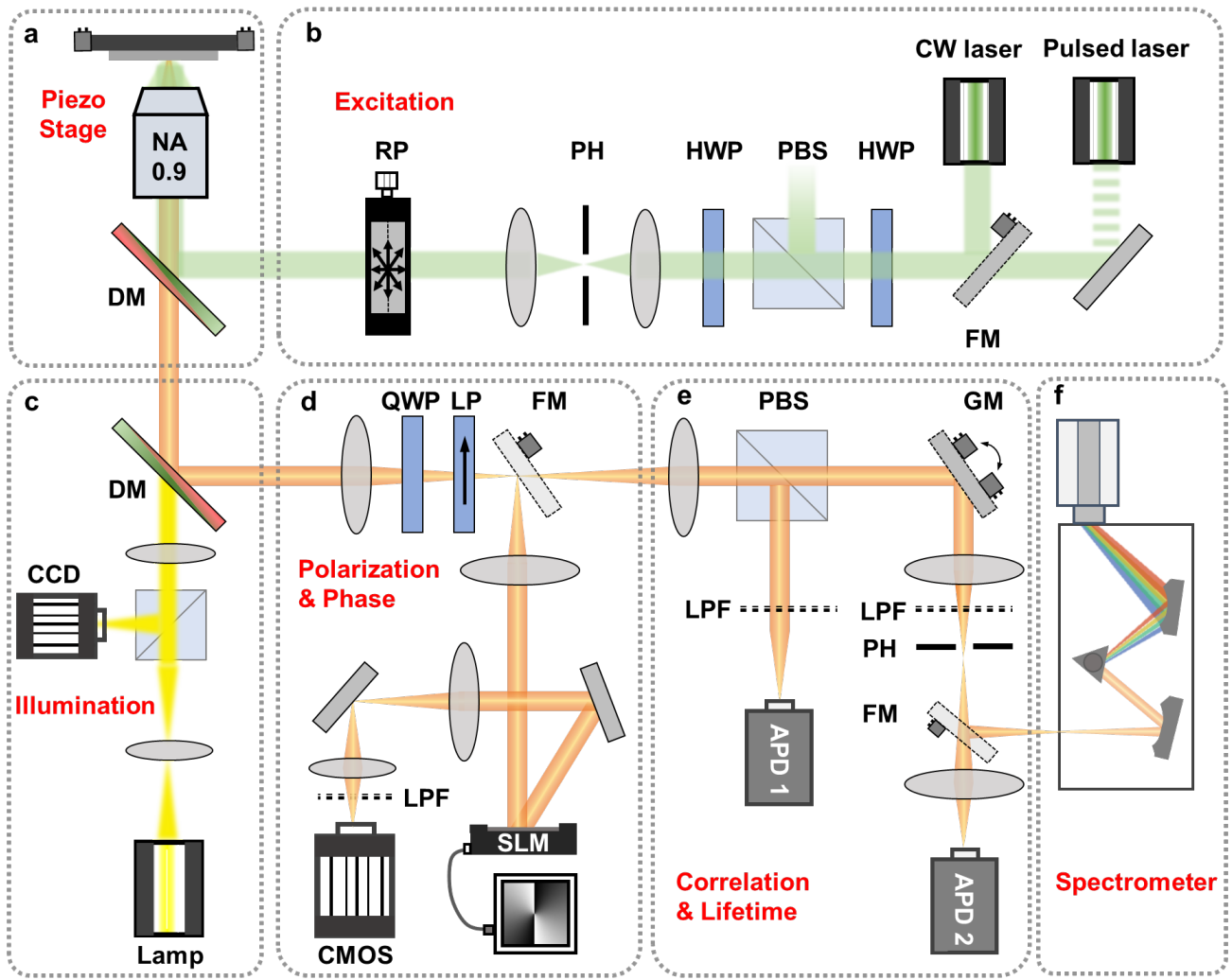

**Supplementary Figure 12 | Schematic of experimental setup. a**, Sample stage. The piezo-stage allows for locating ND-NVs when scanning fluorescence maps. **b**, Incident light for exciting the nanodiamond. **c**, Illumination part for finding the fabricated QE-coupled metasurfaces. **d**, Characterization for polarization states and topological charge. **e**, Characterization for fluorescence image, correlation, decay-rate. Fluorescence photon rate is recorded by avalanche photo diode (APD1), which is filtered from the laser light, by a set of dichroic mirrors (DM) and a long pass filter (LPF). Correlation measurements is recorded by histogramming the timing delay between photon detection events between APD1 and APD2 in a start-stop configuration, using an electronic timing box (Picharp-300; Pico quant). **f**, Characterization for spectrum. CW: continuous wave, RP: Radial Polarization Converter, PBS: polarized beam splitter, PH: pinhole, DM: dichroic mirror, LP: linear polarization, QWP: quarter-wave plate, LPF: 550 nm long pass filter, FM: flip mirror, GM: galvanometric mirror. APD: Avalanche Photodiode.

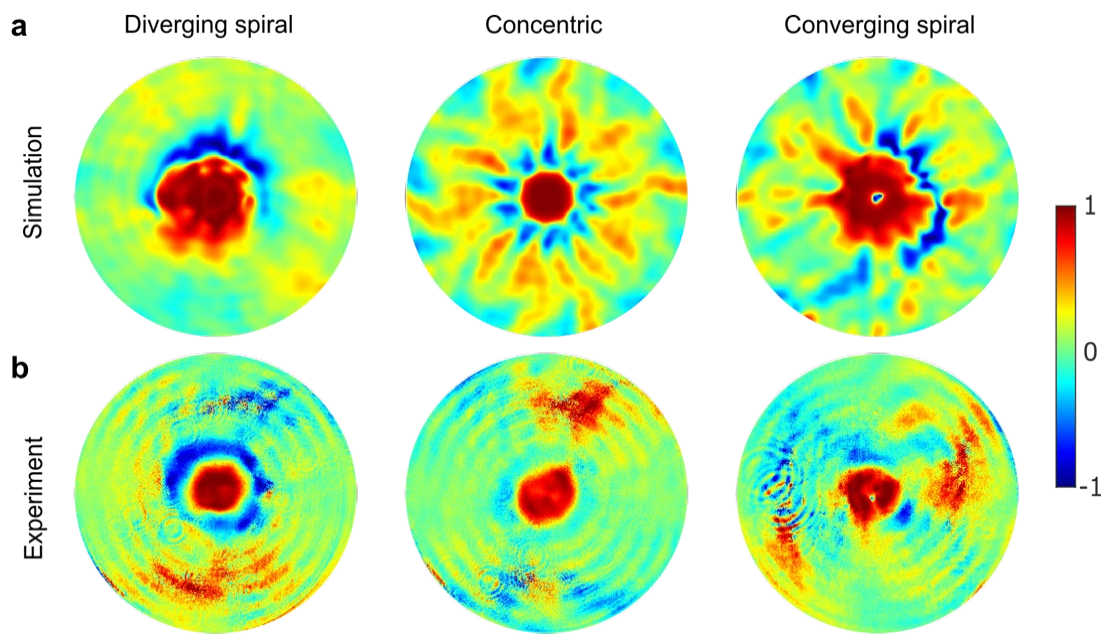

**Supplementary Figure 13 | Characteristics of the circular polarized states. a**, **b**, Simulated (**a**) and measured (**b**) Stokes parameter $S_3$ distributions of the designed diverging, concentric, and converging configurations.

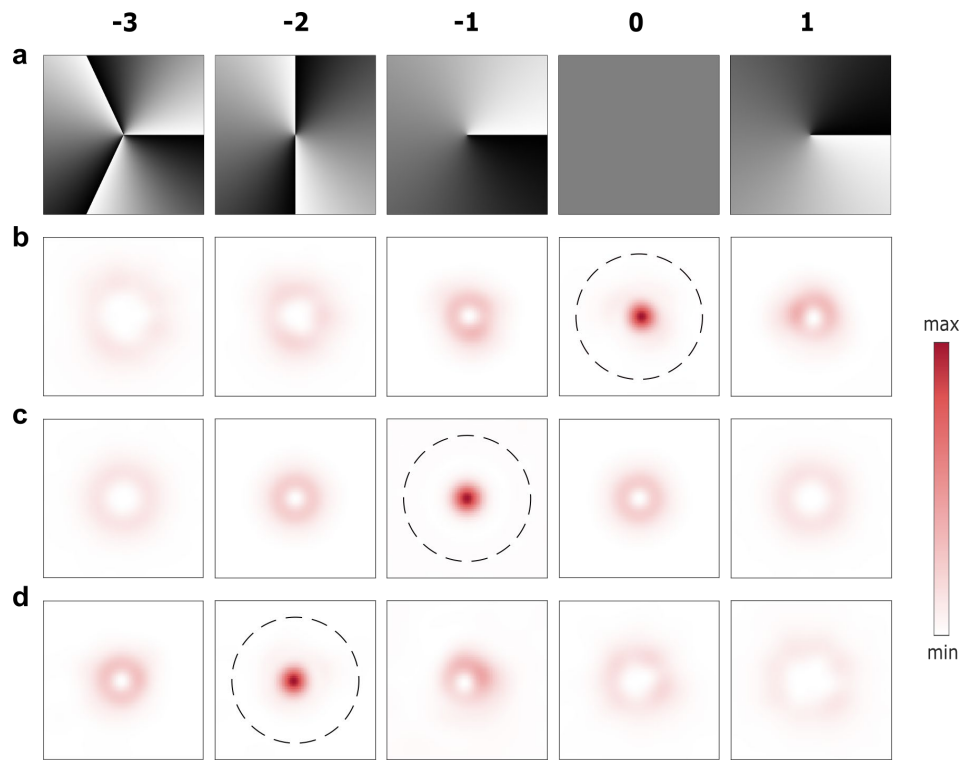

**Supplementary Figure 14 | Simulated vortex states from QE-coupled metasurfaces. a**, Phase distribution for holograms with topological charges with -3, -2, -1, 0, and 1. **b-d**, Simulated intensity distributions of the RCP components of the single-photon emissions that are projected to holograms with (**b**) diverging spiral, (**c**) concentric, (**d**) converging spiral configuration.

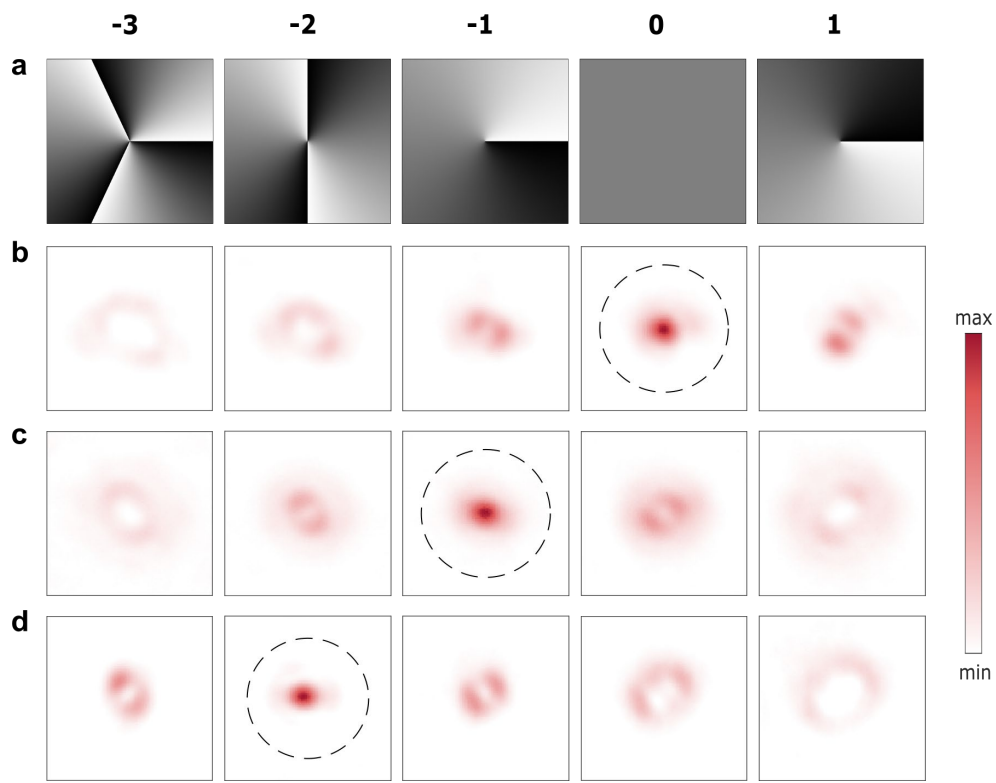

**Supplementary Figure 15 | Measured vortex states of the designed photon sources with ND-NVs. a**, Phase distribution for holograms with topological charges with -3, -2, -1, 0, and 1. **b-d**, Measured intensity distributions of the RCP components of the single-photon emissions that are projected to holograms with (**b**) diverging spiral, (**c**) concentric, (**d**) converging spiral configuration.

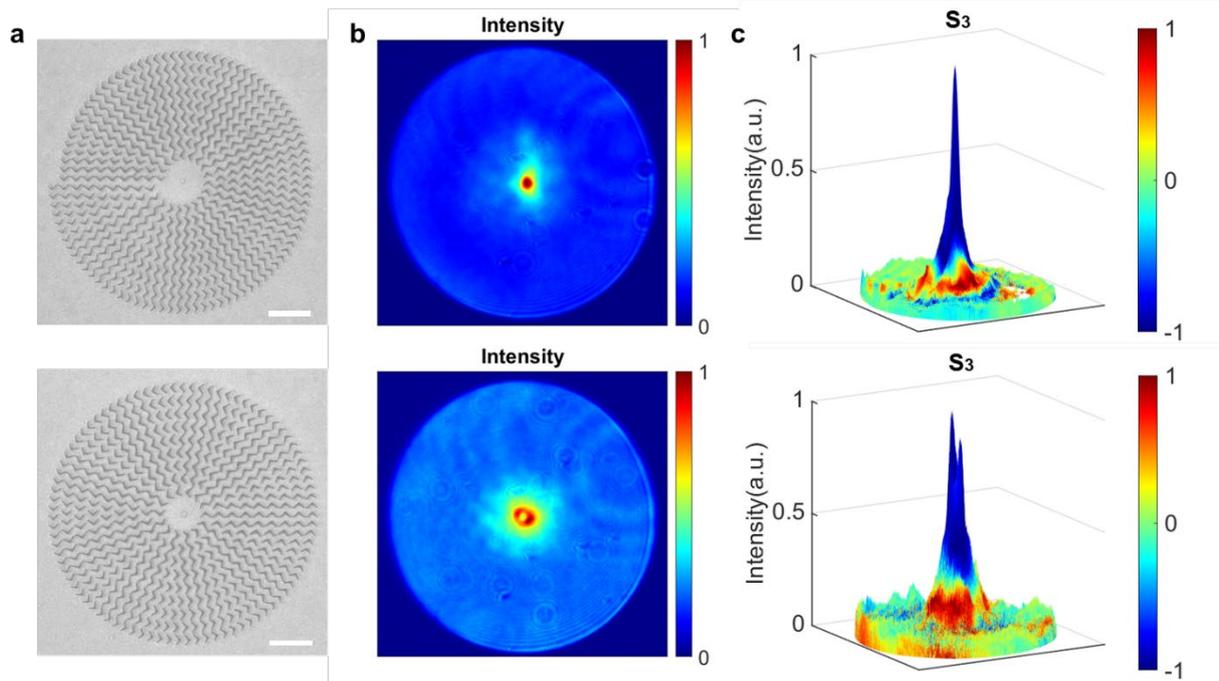

**Supplementary Figure 16 | Experimental demonstration of left-hand circular polarized (LCP) vortex photon sources. a**, SEM images of the fabricated circularly polarized vortex photon sources with topological charge of 0 (top) and 1 (bottom). **b**, Far-field intensity distribution. **c**, 3D representation of the superimposed intensity and polarization distribution. The height indicating the intensity and the colour shows the value of Stokes Vector $S_3$.

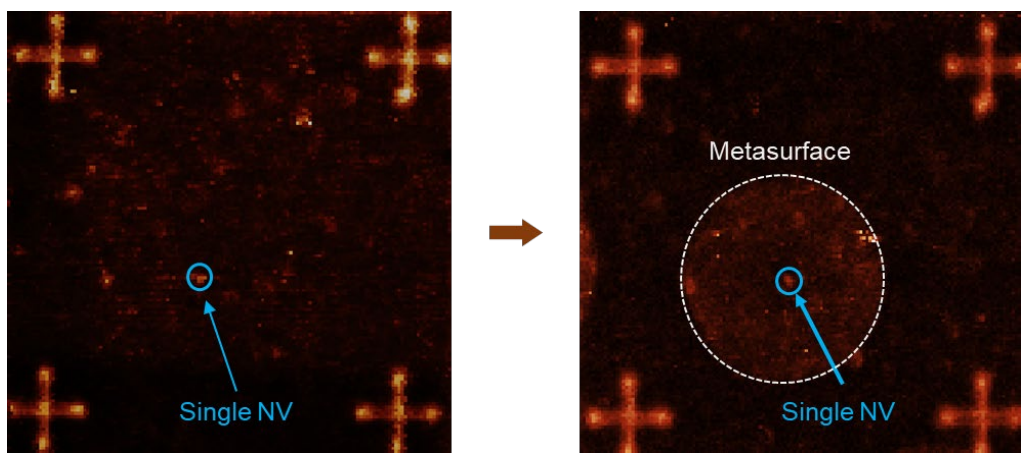

**Supplementary Figure 17 | Fluorescence images showing single-photon sources before and after metasurface fabrication.**

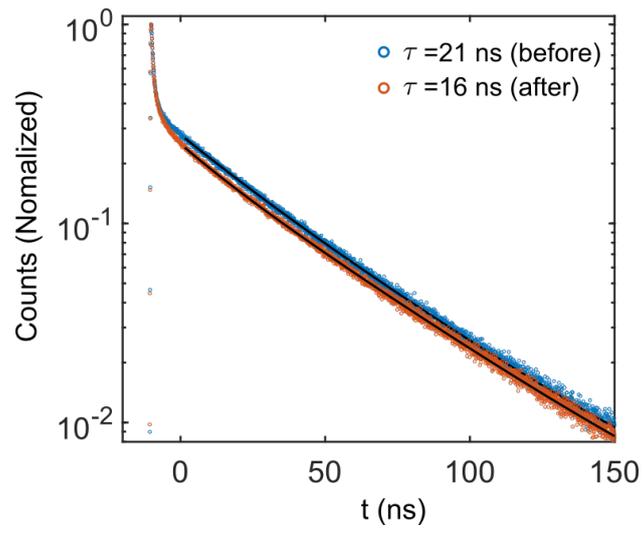

**Supplementary Figure 18 | Lifetime measurements with single photon sources before and after metasurface fabrication.**

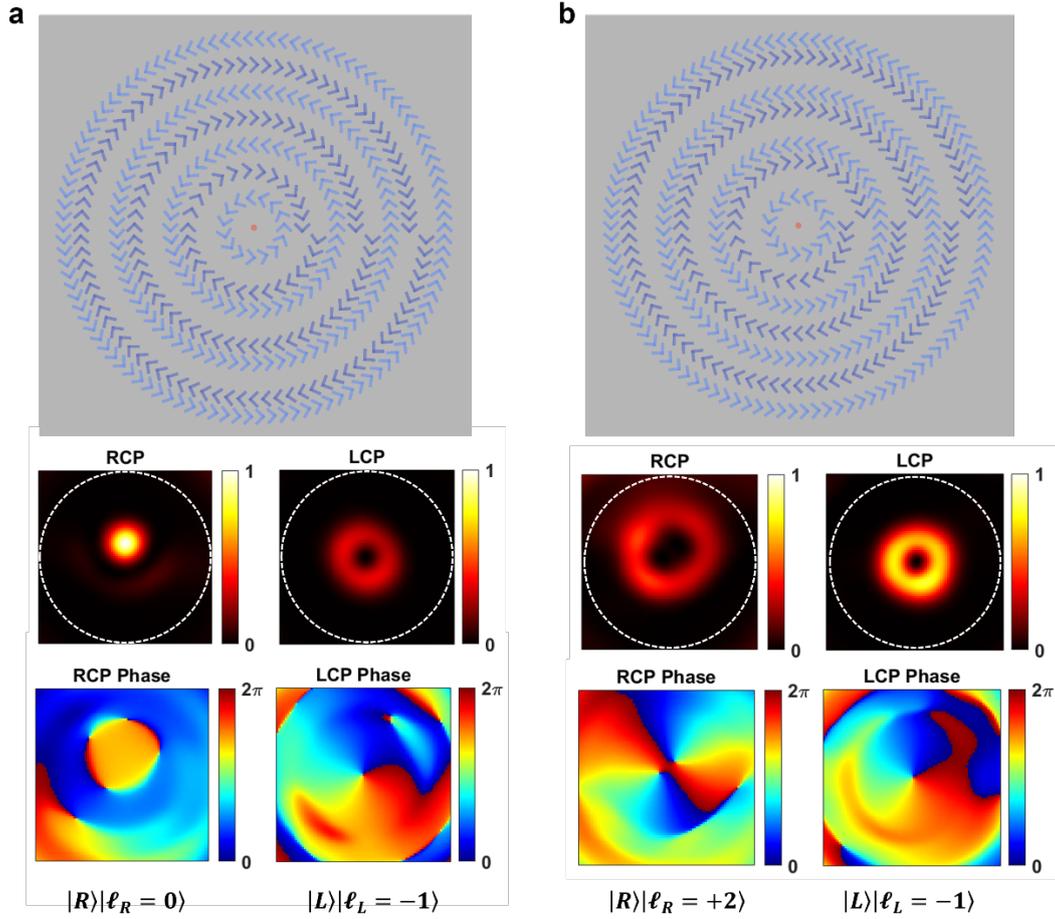

**Supplementary Figure 19 | Interleaved arrangements of QE-coupled quantum metasurfaces. a**, Combination of diverging spiral (RCP element) and concentric (LCP element) for realizing entangled states of $|R\rangle|\ell_R = 0\rangle$ and $|L\rangle|\ell_L = -1\rangle$. **b**, Combination of converging spiral (RCP element) and concentric (LCP element) for realizing entangled states of $|R\rangle|\ell_R = +2\rangle$ and $|L\rangle|\ell_L = -1\rangle$. The first row is the top view of designed QE-coupled quantum metasurfaces, the second row is the RCP and LCP intensities, the third row is the corresponding phases. Note that all the bricks are made by same dielectric materials (HSQ), the difference of colour in first row just identifies the elements with different orientations that contributes to either RCP or LCP. The white dashed circles denote the numerical aperture NA = 0.2.